\documentclass[nofootinbib]{article}

\pdfoutput=1

\usepackage[utf8]{inputenc}

\usepackage{amssymb}

\usepackage{amsmath}

\usepackage{bm}

\usepackage{jcappub}

\usepackage{xcolor}

\allowdisplaybreaks

\newcommand{\beq}{\begin{equation}}
\newcommand{\eeq}{\end{equation}}
\newcommand{\bea}{\begin{eqnarray}}
\newcommand{\eea}{\end{eqnarray}}
\newcommand{\ben}{\begin{enumerate}}
\newcommand{\een}{\end{enumerate}}

\newcommand{\pa}{\partial}

\newcommand{\na}{\nabla}

\newcommand{\ed}{{\rm d}}
\newcommand{\ced}{{\rm D}}
\newcommand{\we}{\wedge}

\newcommand{\ti}{\tilde}

\renewcommand\({\left(}
\renewcommand\){\right)}
\renewcommand\[{\left[}
\renewcommand\]{\right]}

\newcommand{\nn}{\nonumber}

\newcommand{\al}{\alpha}
\newcommand{\be}{\beta}
\newcommand{\ga}{\gamma}
\newcommand{\Ga}{\Gamma}
\newcommand{\de}{\delta}
\newcommand{\De}{\Delta}
\newcommand{\ep}{\epsilon}

\newcommand{\et}{\eta}

\newcommand{\Te}{\Theta}

\newcommand{\ka}{\kappa}
\newcommand{\la}{\lambda}
\newcommand{\La}{\Lambda}

\newcommand{\si}{\sigma}

\newcommand{\ta}{\tau}
\newcommand{\ph}{\phi}
\newcommand{\vph}{\varphi}
\newcommand{\ch}{\chi}

\newcommand{\Om}{\Omega}

\newcommand{\cH}{{\cal H}}

\title{General Relativistic Cosmological N-body Simulations I: time integration}

\author[a]{David Daverio,}
\author[b]{Yves Dirian,}
\author[b]{Ermis Mitsou}

\affiliation[a]{Centre for Theoretical Cosmology, Department of Applied Mathematics and Theoretical Physics, \\ University of Cambridge, Wilberforce Road, Cambridge CB3 0WA, United  Kingdom}
\affiliation[b]{Center for Theoretical Astrophysics and Cosmology, Institute for Computational Science, \\ University of Z\"urich, CH--8057 Z\"urich, Switzerland}

\emailAdd{dd415@damtp.cam.ac.uk}
\emailAdd{yves.dirian@ics.uzh.ch}
\emailAdd{ermitsou@physik.uzh.ch}

\abstract{This is the first in a series of papers devoted to fully general-relativistic $N$-body simulations applied to late-time cosmology. The purpose of this paper is to present the combination of a numerical relativity scheme, discretization method and time-integration algorithm that provides satisfyingly stable evolution. More precisely, we show that it is able to pass a robustness test and to follow scalar linear modes around an expanding homogeneous and isotropic space-time. Most importantly, it is able to evolve typical cosmological initial conditions on comoving scales down to tenths of megaparsecs with controlled constraint and energy-momentum conservation violations all the way down to the regime of strong inhomogeneity.}

\begin{document}

\maketitle

\flushbottom

\section{Introduction}

The forthcoming advances in the observations of the cosmological large scale structure (LSS) \cite{LSST,EUCLID,DESI,SKA,WFIRST} require a proportionate refinement of our theoretical predictions, not only to exploit the increased amount and precision of the data, but also in order to correctly interpret them. The standard numerical approach to study the non-linear LSS dynamics is the Newtonian $N$-body simulation \cite{Teyssier:2001cp, Springel:2005mi, Potter:2016ttn}, which essentially emulates the Boltzmann equation for ``cold" collisionless matter in Newtonian gravity. Such simulations ignore the relativistic effects of General Relativity (GR) in the dynamics, but also in the reconstruction of observables, since they do not take into account the full geometrical information of space-time. The Newtonian approximation only applies to cosmological models where matter is non-relativistic and effectively decoupled from relativistic degrees of freedom, such as in $\La$CDM, thus excluding several alternative descriptions of the dark sector. Moreover, it also fails at scales comparable to the Hubble radius, which the forthcoming missions will be able to probe. At such scales the causality imposed by relativity can no longer be ignored and relativistic effects are known to become important, at least in the observables \cite{Yoo:2009au, Yoo:2010ni, Bonvin:2011bg, Jeong:2011as, Bruni:2011ta, Challinor:2011bk, Yoo:2012se, Bonvin:2014owa, Bartolo:2015qva, Alonso:2015uua, Alonso:2015sfa, Umeh:2016nuh, Jolicoeur:2017nyt, Biern:2017bzo, Tansella:2017rpi, Jolicoeur:2017eyi, Bonvin:2017req, Bonvin:2018ckp, Tansella:2018sld, Scaccabarozzi:2018vux, Jolicoeur:2018blf, Castiblanco:2018qsd}. 

Nevertheless, these limitations can be circumvented to some extent with the help of analytical tools that were developed in the last decade. There are now refined perturbative expansions of the Einstein equations around the Friedmann-Lema\^itre-Robertson-Walker solution (FLRW) that are able to capture the non-linear matter dynamics \cite{Green:2010qy, Brustein:2011dy, Kopp:2013tqa, Milillo:2015cva, Rampf:2016wom, Eingorn:2015hza, Eingorn:2016kdt, Brilenkov:2017gro, Goldberg:2016lcq, Goldberg:2017gsm}. Along with mapping techniques or appropriate gauge choices, one can then use Newtonian $N$-body simulations to effectively solve the non-linear dynamics of the relativistic theory \cite{Chisari:2011iq, Green:2011wc, Rampf:2012pu, Rampf:2013ewa, Rampf:2013dxa, Bruni:2013mua, Rigopoulos:2013nda, Rampf:2014mga, Rigopoulos:2014rqa, Fidler:2015npa, Christopherson:2015ank, Hahn:2016roq, Fidler:2016tir, Brandbyge:2016raj, Borzyszkowski:2017ayl, Fidler:2017pnb, Fidler:2017ebh, Adamek:2017grt, Fidler:2018bkg, Tram:2018znz, Dakin:2019dxu, Dakin:2019vnj, Adamek:2019aad}. In \cite{Adamek:2013wja, Adamek:2014xba, Adamek:2015eda, Adamek:2016zes, Adamek:2017uiq} the authors went one step further by developing the first $N$-body code based on such a truncation of the Einstein equations, thus including all the information of the metric tensor and capturing the dominant relativistic effects.

\begin{figure}[htbp]
\includegraphics[width=0.3333\columnwidth]{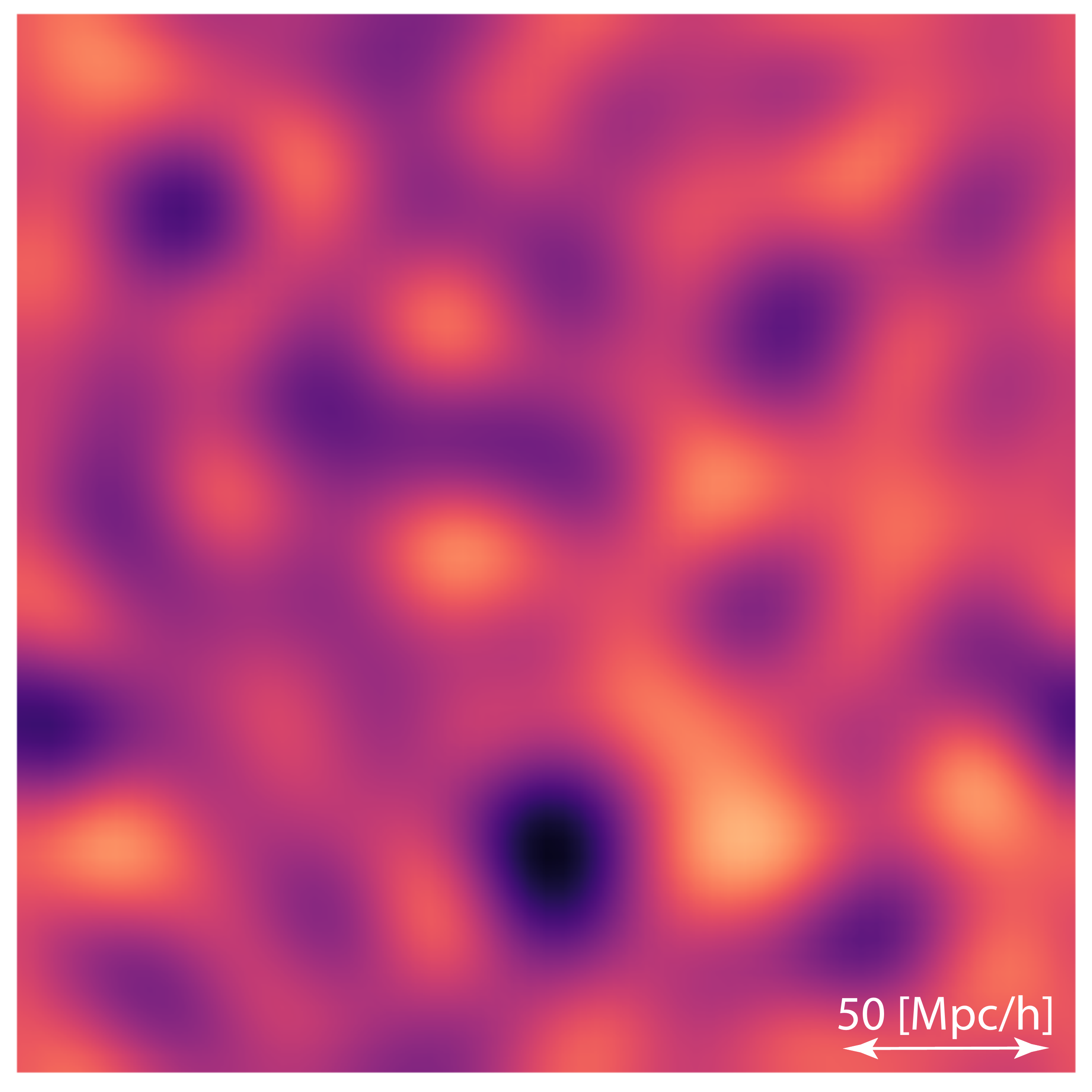} 
\includegraphics[width=0.3333\columnwidth]{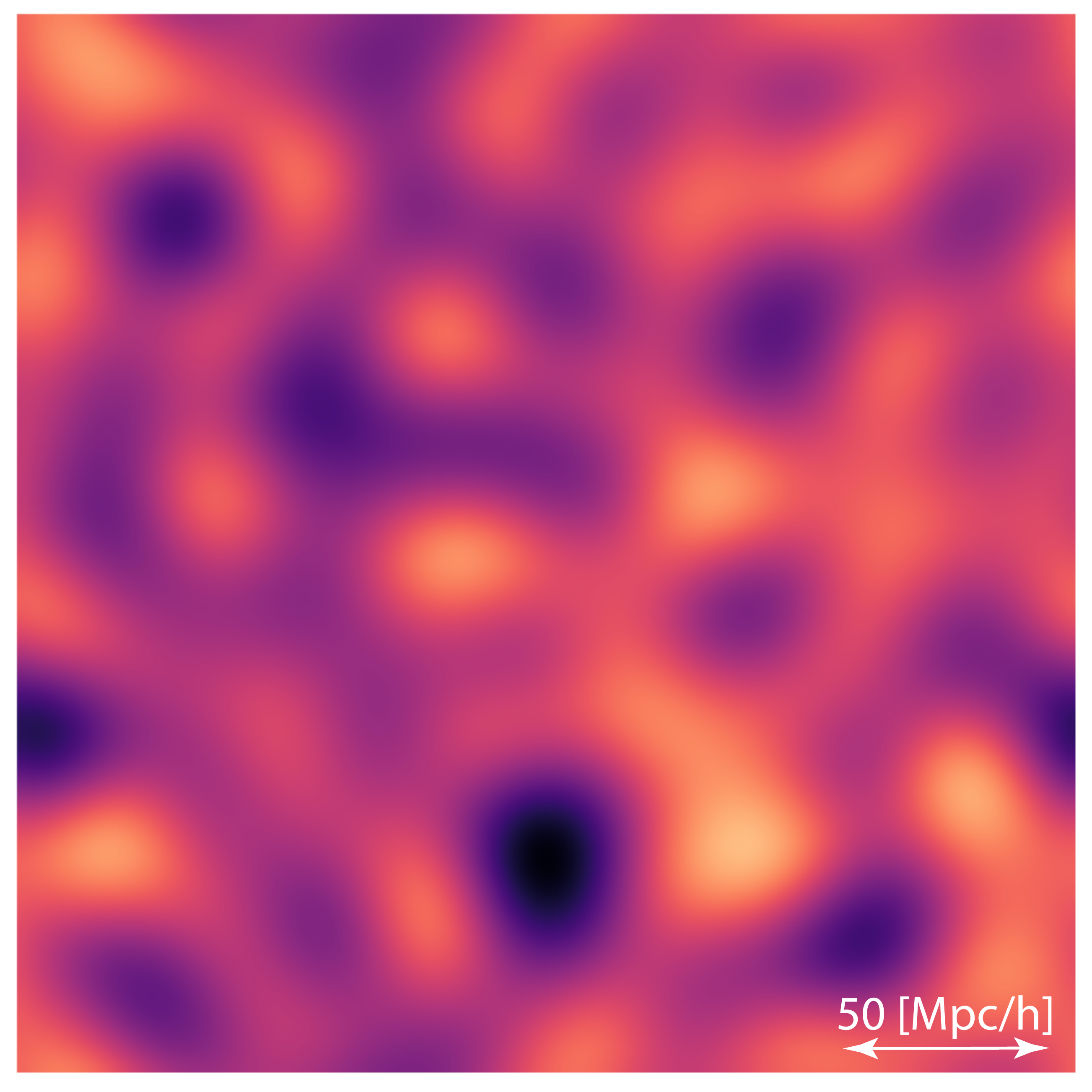} 
\includegraphics[width=0.3333\columnwidth]{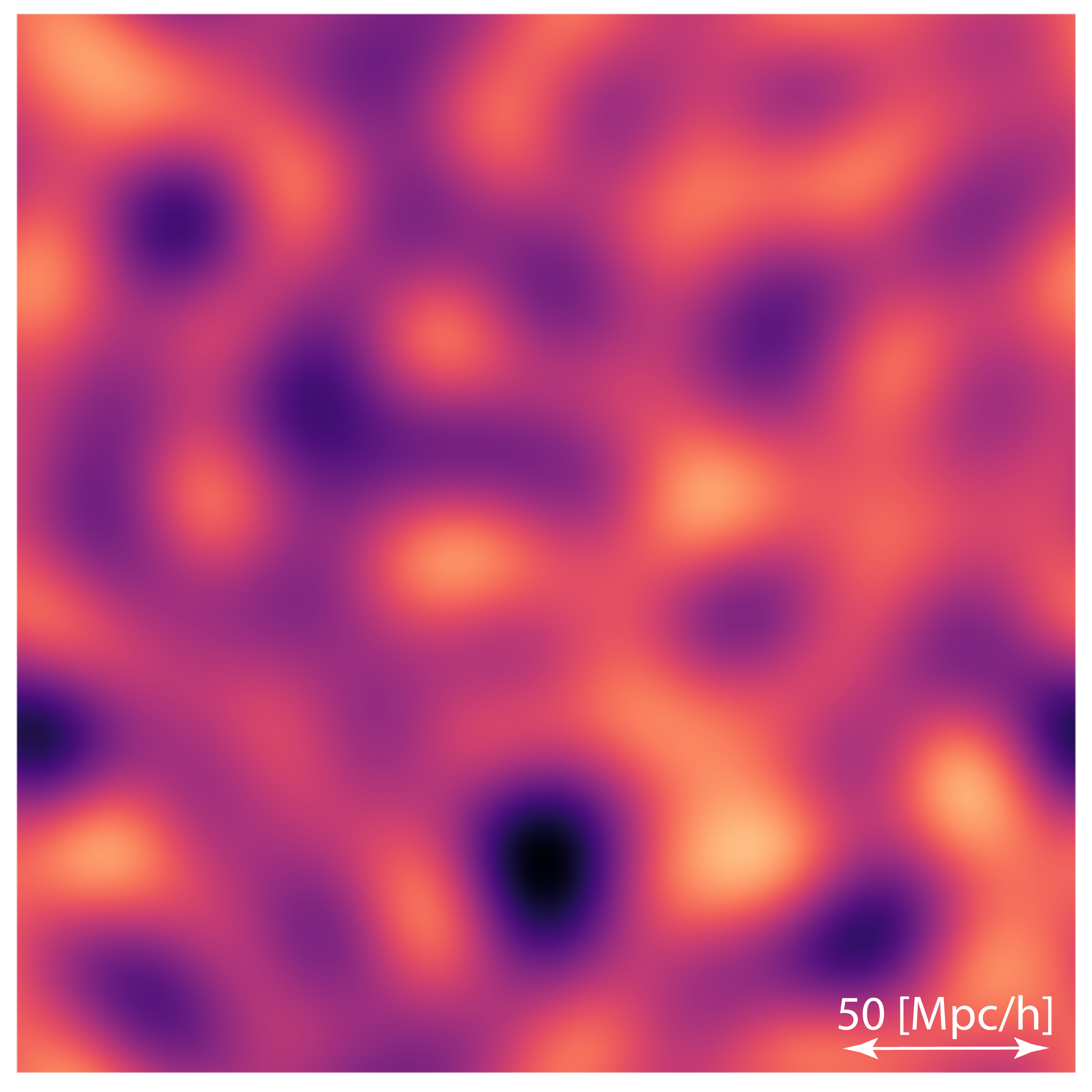} 
\caption{Two-dimensional slice of the energy density $E$ at redshift 100 with spatial resolution $\De x = 4$ (left), $\De x = 2$ (center) and $\De x = 1$ (right).}
\label{fig:z100}
\end{figure}

\begin{figure}[htbp]
\includegraphics[width=0.3333\columnwidth]{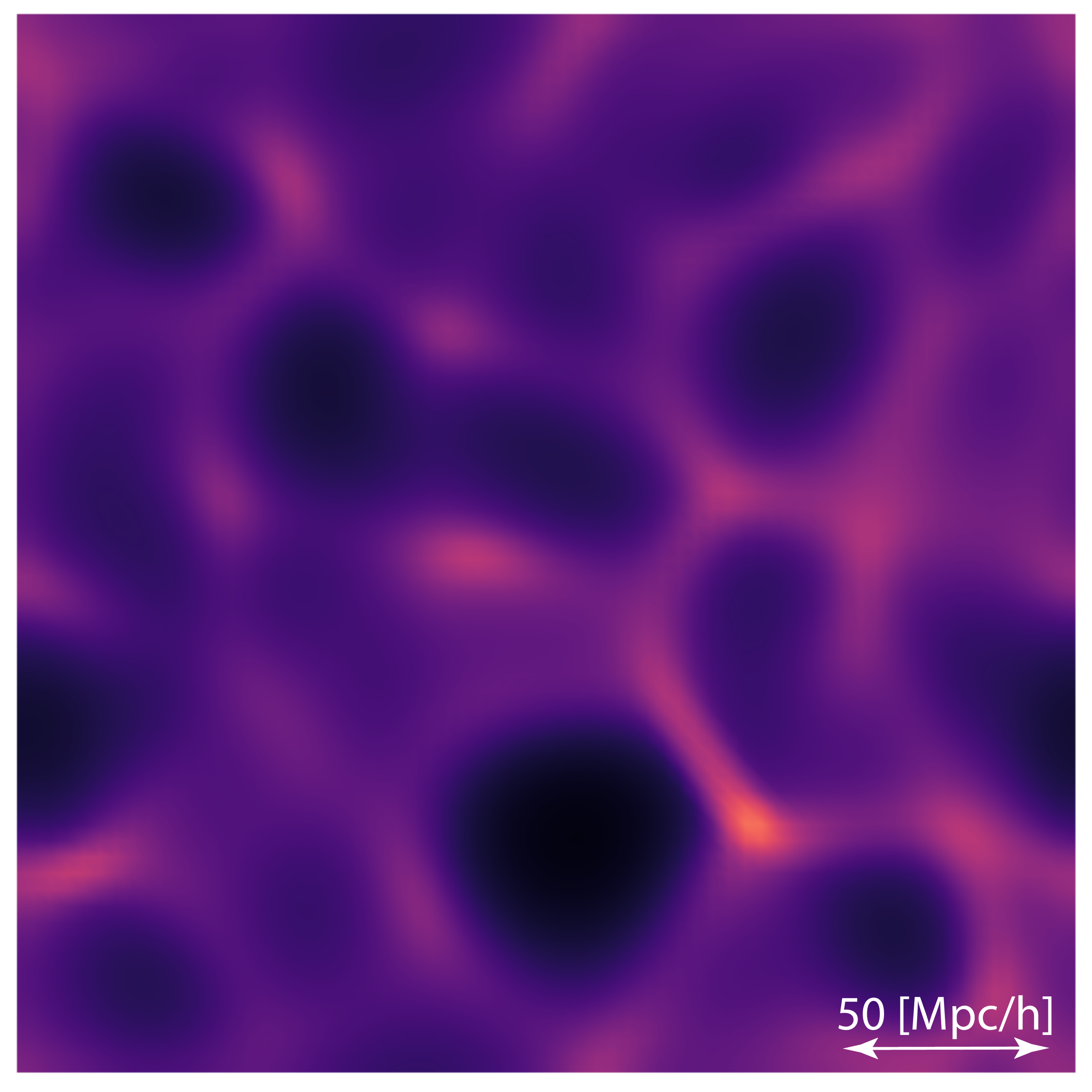} 
\includegraphics[width=0.3333\columnwidth]{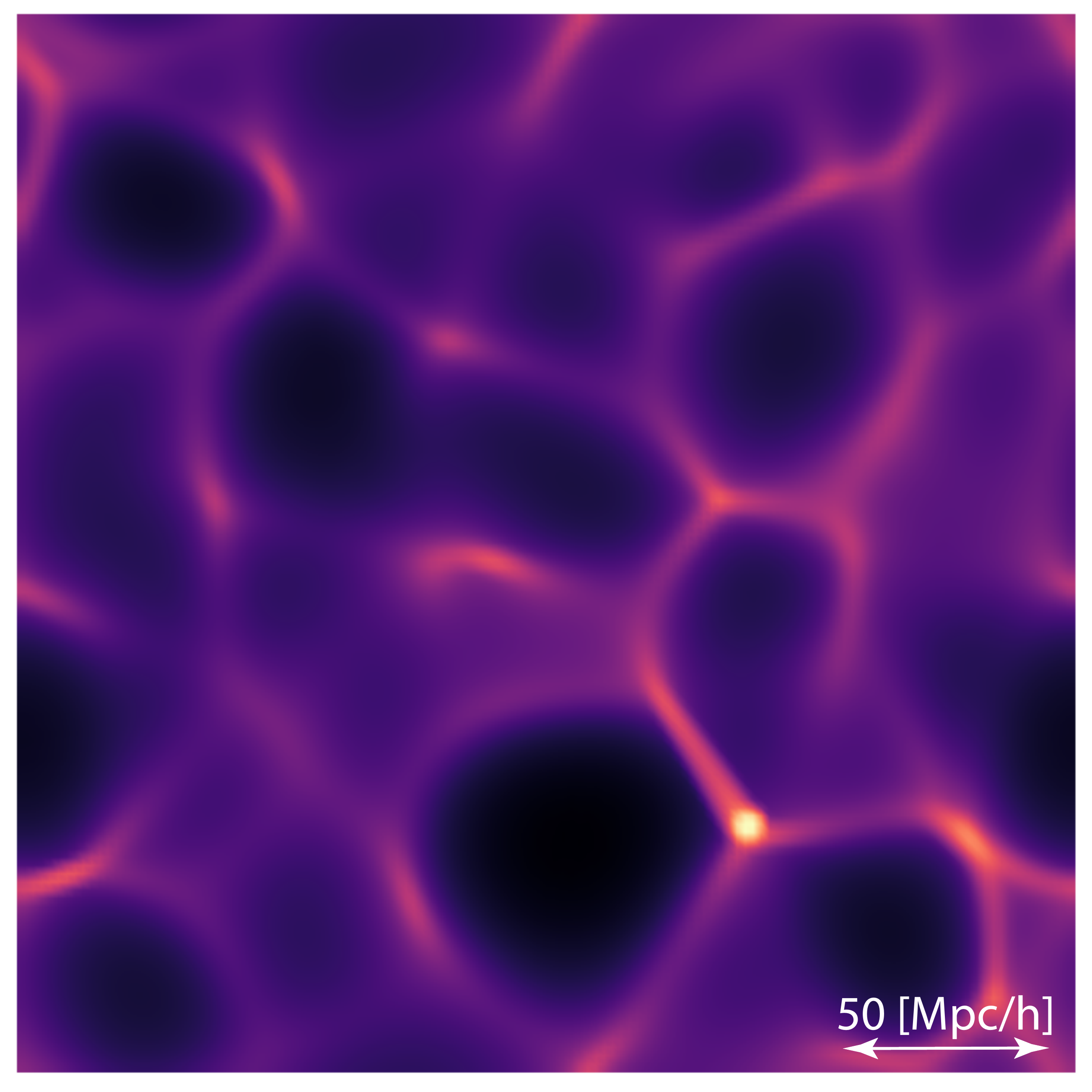} 
\includegraphics[width=0.3333\columnwidth]{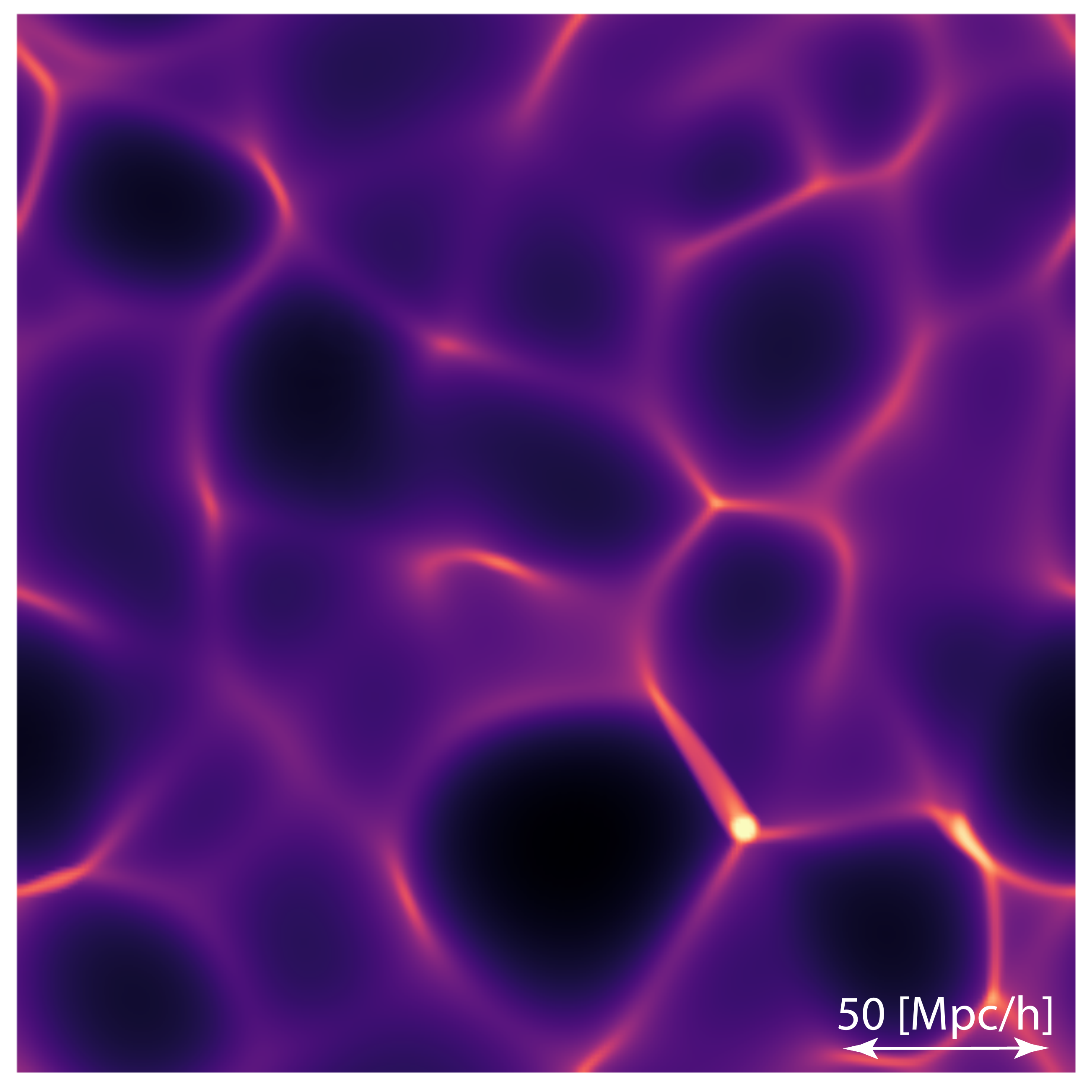} 
\caption{Two-dimensional slice of the energy density $E$ at redshift 0 with spatial resolution $\De x = 4$ (left), $\De x = 2$ (center) and $\De x = 1$ (right). }
\label{fig:z0}
\end{figure}

Although the above methods are certainly very convenient, they are still defined within a perturbative approach. Some argue\footnote{See \cite{Buchert:2007ik, Buchert:2011yu, Clarkson:2011zq, Ellis:2011hk, Rasanen:2011ki, Wiltshire:2011vy, Buchert:2011sx, Buchert:2015iva} for reviews and discussions on the issue of  ``backreaction" of small scale inhomogeneities on the large scale dynamics, see \cite{Baumann:2010tm, Green:2010qy, Green:2011wc, Green:2014aga, Green:2015bma, Green:2016cwo} for counter-arguments and \cite{Adamek:2014gva, Adamek:2017mzb, Adamek:2018rru, Giblin:2015vwq, Bentivegna:2015flc, Giblin:2016mjp, East:2017qmk, Macpherson:2018btl, Macpherson:2018akp} for related numerical investigations.} that, in the presence of strong inhomogeneity, there could be important non-perturbative effects invalidating any perturbative treatment. This has motivated a recent interest in simulations that solve the fully non-linear Einstein equations \cite{Giblin:2015vwq, Bentivegna:2015flc, Mertens:2015ttp, Bentivegna:2016stg, Giblin:2016mjp, Daverio:2016hqi, Macpherson:2016ict, Giblin:2017juu, East:2017qmk, Macpherson:2018btl, Macpherson:2018akp}, i.e. the application of numerical relativity (NR) to cosmology. In these cases, however, the matter sector has always been modeled as a pressureless perfect fluid field in a grid-based approach. Consequently, it cannot describe the correct (collisionless) dynamics at scales where shell-crossing occurs, which roughly coincides with the scales at which the dynamics become non-linear. One way of describing a cosmology with ``granular" matter within NR, which has also received particular focus, are simulations of lattice black hole configurations \cite{Clifton:2009jw, Clifton:2012qh, Yoo:2012jz, Bruneton:2012cg, Bentivegna:2012ei, Bruneton:2012ru, Bentivegna:2013jta, Bentivegna:2018koh, Giblin:2019pql}, but the high degree of symmetry makes such solutions too idealized to describe realistic dark matter dynamics.

The status quo naturally leads us to consider the potential advantages of an $N$-body NR approach, i.e. taking into account all non-linear and relativistic effects, while solving the correct matter dynamics at small scales. On the one hand, such simulations would serve as a control reference for comparing with approximative methods, both analytical and numerical, thus testing their robustness and potentially settling issues if ambiguous results arise between different approaches. On the other hand, if any non-perturbative effects turn out to occur, be it those that have been speculated over or genuinely new ones, such codes would be the only way to capture them. 

The first $N$-body NR simulations have been performed in the context of gravitational collapse dynamics in the mid-eighties \cite{ST1, ST2, ST3, ST4, RST, ST5} for configurations of reduced dimensionality, while the first studies of three-dimensional configurations occurred in the late nineties \cite{Shibata:1999va, Shibata:1999wi}, with a recent revival in \cite{Yamada:2011br, Yoo:2016kzu, East:2019bwu}. In \cite{Pretorius:2018lfb} it is the case of massless particles that was considered to study the collision of plane-fronted gravitational waves. The only appearance of such simulations applied to cosmology is, to our knowledge, in \cite{Pretorius:2018lfb,East:2019chx,Giblin:2018ndw}. However, the most complicated configuration considered in these papers is the triple-mode inhomogeneity around the FLRW space-time with equal comoving wavelengths, i.e. $\de(\vec{x}) = \sum_{i=1}^3 A_i \sin(k x^i)$. In \cite{Pretorius:2018lfb,East:2019chx} the authors consider the case $A_i \sim 10^{-3} - 10^{-2}$ and a comoving wavelength that is four times the initial comoving Hubble radius. In \cite{Giblin:2018ndw}, where the aim is to study quantitatively the deviation from linear cosmological perturbation theory, the authors consider comoving wavelengths of 400 and 100 Mpc with an amplitude $A$ corresponding to the typical power at these scales. Finally, let us also mention another recent $N$-body code for cosmological simulations \cite{Barrera-Hinojosa:2019mzo} which employs the so called ``fully constrained formulation" of GR \cite{Bonazzola:2003dm,CorderoCarrion:2008cx,CorderoCarrion:2008nf} with an approximation which essentially neglects its propagating degrees of freedom (gravitational waves). This reduces the problem to a set of non-linear elliptic field equations, as in the case of Newtonian $N$-body simulations.

Our aim is to develop a combination of numerical methods that will ultimately allow one to perform realistic three-dimensional cosmological $N$-body NR simulations. This endeavor presents new computational challenges compared to Newtonian $N$-body and grid-based NR codes, which we will address in a series of papers. The present paper focuses on the time evolution of the system and its stability, providing in particular the numerical relativity scheme, discretization method and time-integration algorithm. We show that this combination is able to solve the FLRW solution robustly, i.e. it is stable under the injection of white noise, and it can accurately follow a scalar linear mode fluctuation around that solution. The most important result is that it is able to evolve typical cosmological initial configurations for a pure dark matter universe, resolving comoving scales down to tenths of megaparsecs (see figures \ref{fig:z100} and \ref{fig:z0}), with a controlled violation of the constraints and of the energy-momentum conservation. In particular, the Hamiltonian and momentum constraints are satisfied with an average relative precision of $\sim 10^{-6}$ and $\sim 10^{-2}$, respectively, while the energy-momentum conservation equations are satisfied at the level of $\sim 10^{-2}$, all the way down to the regime of strong inhomogeneity $z \sim 0$. At large enough scales, where the linear theory holds, we also follow the matter power spectrum with a relative precision of $\sim 10^{-2}$. The convergence is of the expected order, but not after redshift $z \sim 10$, and we believe that adaptive resolution, in both the mesh and the phase space sampling, will be able to resolve this issue. 

In section \ref{sec:nummeth} we present the involved numerical methods, in section \ref{sec:tests} we present the results of the aforementioned tests and in section \ref{sec:conclusion} we conclude. All the required equations related to our tests are derived in the appendices. We work in the following units 
\beq
8\pi G = c = {\rm Mpc}/h = 1 \, ,
\eeq
and all the present tests are performed with zero cosmological constant. Our code is implemented on top of the {\sc lat}field{\sc 2} library \cite{David:2015eya}.

\section{Numerical methods} \label{sec:nummeth}

\subsection{Evolution equations}

On the gravitational side we consider the damped CCZ4 scheme \cite{Bona:2003fj, Gundlach:2005eh, Alic:2011gg}, which can be seen as a generalization of the BSSNOK scheme \cite{NOK, Shibata:1995we, Baumgarte:1998te} involving an additional ``pure-constraint" field $\Te$ that helps diluting constraint violation by propagating it away. We base our choice of scheme on a recent ``Apples-with-Apples" comparison \cite{Alcubierre:2003pc, Babiuc:2007vr} of CCZ4 with closely related ones that we performed in \cite{Daverio:2018tjf}, and also the tests performed in this paper. The evolution equations are
\bea
\ced_t \Te & = & \al \[ - K \Te - \ka_1 \( 2 + \ka_2 \) \Te - H \] - \chi \ti{Z}^i \pa_i \al   \, , \label{eq:TeEvol} \\
\ced_t \hat{\Ga}^i & = & 2\al \[ \ti{\Ga}^i_{jk} \ti{A}^{jk} - \frac{3}{2}\, \ti{A}^{ij} \chi^{-1} \pa_j \chi + \ti{\ga}^{ij} \( \pa_j \Te - \frac{2}{3}\, \pa_j K  - P_j \) - \( \frac{2}{3} \, K + \ka_1 \) \ti{Z}^i \] \nn \\
 & & - 2 \Te \ti{\ga}^{ij} \pa_j \al - 2 \ti{A}^{ij} \pa_j \al + \ti{\ga}^{jk} \pa_j \pa_k \be^i + \frac{1}{3}\, \ti{\ga}^{ij} \pa_j \pa_k \be^k  - \hat{\Ga}^j \pa_j \be^i + \frac{2}{3}\, \hat{\Ga}^i \pa_j \be^j \, , \\
\ced_t \chi & = & \frac{2}{3}\, \chi \[ \al K - \pa_i \be^i \] \, , \\
\ced_t K & = & \al \[ K^2 - 2  K \Te - 3 \ka_1 \( 1 + \ka_2 \) \Te + 2 \ti{Z}^i \pa_i \chi - \frac{3}{2}\, E  \] + \chi \ti{\Ga}^i \pa_i \al \nn \\
 & & +\, \ti{\ga}^{ij} \[ - \chi \pa_i \pa_j \al + \frac{1}{2}\, \pa_i \chi \pa_j \al + \al \( \hat{R}_{ij} + \frac{1}{2}\, \chi S_{ij} \) \] \, , \\
\ced_t \ti{\ga}_{ij} & = & - 2 \al \ti{A}_{ij} + \ti{\ga}_{ik} \pa_j \be^k + \ti{\ga}_{jk} \pa_i \be^k - \frac{2}{3}\, \ti{\ga}_{ij} \pa_k \be^k \, , \\
\ced_t \ti{A}_{ij} & = & \al \[ - 2 \ti{\ga}^{kl} \ti{A}_{ik} \ti{A}_{jl} + \( K - 2 \Te \) \ti{A}_{ij} \] \nn \\
 & & + \[ \chi \( - \pa_i \pa_j \al + \ti{\Ga}^k_{ij} \pa_k \al \) - \pa_{(i} \chi \pa_{j)} \al + 2 \al \ti{Z}^k \ti{\ga}_{k(i} \pa_{j)} \chi + \al \( \hat{R}_{ij} - \chi S_{ij} \)  \]^{\rm TF}  \nn \\
 & & +\,  \ti{A}_{ik} \pa_j \be^k + \ti{A}_{jk} \pa_i \be^k - \frac{2}{3}\, \ti{A}_{ij} \pa_k \be^k \, , 
\eea
where
\bea
\ced_t & := & \pa_t - \be^i \pa_i \, , \\
\ti{\Ga}^i & := & - \pa_j \ti{\ga}^{ij} \, , \\
\ti{Z}^i & := & \frac{1}{2} \[ \hat{\Ga}^i - \ti{\Ga}^i \] \, , \\
\ti{\Ga}_{kij} & := & \frac{1}{2} \( \pa_i \ti{\ga}_{jk} + \pa_j \ti{\ga}_{ik} - \pa_k \ti{\ga}_{ij} \) \, , \\
\ti{\Ga}^k_{ij} & := & \ti{\ga}^{kl} \ti{\Ga}_{lij} \, , \\
\hat{R}_{ij} & := & \chi \[ - \frac{1}{2}\, \ti{\ga}^{kl} \pa_k \pa_l \ti{\ga}_{ij} + \ti{\ga}_{k(i} \pa_{j)} \hat{\Ga}^k + \ti{\Ga}_{(ij)k} \hat{\Ga}^k + \ti{\ga}^{kl} \( \ti{\Ga}^m_{ki} \ti{\Ga}_{mlj} + 2\ti{\Ga}^m_{k(i} \ti{\Ga}_{j)ml} \) \] \label{eq:Rhat}  \\
 & & +\, \frac{1}{2} \[ \pa_i \pa_j \chi - \frac{1}{2}\, \chi^{-1} \pa_i \chi \pa_j \chi + \ti{\ga}_{ij} \ti{\ga}^{kl} \( \pa_k \pa_l \chi - \frac{3}{2}\, \chi^{-1} \pa_k \chi \pa_l \chi \) - \ti{\Ga}^k_{ij} \pa_k \chi - \ti{\ga}_{ij} \hat{\Ga}^k \pa_k \chi  \] \, , \nn
\eea
and are subject to the constraint equations
\bea
D & := & \det \ti{\ga} - 1 = 0 \, , \label{eq:Dconstraint} \\
D' & := & \ti{\ga}^{ij} \ti{A}_{ij} = 0 \, , \label{eq:Tconstraint} \\  
\Te & = & 0 \, , \\
\ti{Z}^i & = & 0 \, , \\
H & := & E - \frac{1}{3}\, K^2 + \frac{1}{2}\, \ti{A}_{ij} \ti{A}^{ij} - \frac{1}{2}\, \ti{\ga}^{ij} \hat{R}_{ij} - \ti{Z}^i \pa_i \chi = 0 \, , \label{eq:H} \\
M_i & := & P_i - \ti{\ga}^{jk} \[ \pa_j \ti{A}_{ki} - \ti{A}_{li} \ti{\Ga}^l_{kj} - \ti{A}_{kl} \ti{\Ga}^l_{ij} - \frac{3}{2}\, \ti{A}_{ij} \chi^{-1} \pa_k \chi \] + \frac{2}{3}\, \pa_i K = 0 \, . \label{eq:Mi}
\eea
All indices are displaced using the conformal 3-metric $\ti{\ga}_{ij}$. The line-element reads
\beq
\ed s^2 = - \al^2 \ed t^2 + \ga_{ij} \( \ed x^i + \be^i \ed t \) \( \ed x^j + \be^j \ed t \) \, ,
\eeq
where $\al$ is the lapse function, $\be^i$ is the shift vector, 
\beq
\ga_{ij} := \chi^{-1} \ti{\ga}_{ij} \, , \hspace{1cm} K_{ij} := \chi^{-1} \( \ti{A}_{ij} + \frac{1}{3}\, \ti{\ga}_{ij} K \) \, , 
\eeq
are the 3-metric and extrinsic curvature of the $t = {\rm const.}$ hypersurfaces, while $E$, $P_i$ and $S_{ij}$ are the energy, momentum and stress densities in the canonical frame $n = \al^{-1} \( \pa_t - \be^i \pa_i \)$. By redefining $K \to K +2 \Te$ one obtains the equally well-performing Z4cc scheme considered in \cite{Daverio:2018tjf}, which is then related to BSSNOK by simply setting $\Te = 0$.\footnote{By setting to zero some of the pure-constraint terms of Z4cc, one obtains the Z4c scheme proposed in \cite{Bernuzzi:2009ex,Weyhausen:2011cg} and tested in \cite{Cao:2011fu}, which we included in our comparison \cite{Daverio:2018tjf}.} We refer the reader to \cite{Daverio:2018tjf} for a derivation of the above equations. 

In this paper we will consider the following Bona-Mas\'o slicing \cite{Bona:1994dr,Alcubierre:1138167}
\beq \label{eq:BMal}
\ced_t \al = - \frac{1}{3}\, \al^2 \[ K - 2 \Te \] \, ,
\eeq
and zero shift vector
\beq \label{eq:0shift}
\be^i = 0 \, .
\eeq
This slicing choice corresponds to the conformal time parametrization on the FLRW solution, which we will keep denoting by ``$t$" contrary to the usual convention in cosmology. For the analytical solution where $\Te = 0$, equation \eqref{eq:BMal} can be solved analytically
\beq \label{eq:alsol}
\al = Q \chi^{-1/2} \, ,
\eeq 
where $Q \equiv Q(\vec{x})$ is an arbitrary space field and corresponds to the residual gauge freedom of choosing the initial conditions of $\al$.

The parameters $\ka_1$ and $\ka_2$, introduced in \cite{Gundlach:2005eh}, are free to choose and can be space-time dependent, since they multiply ``pure-constraint" terms. The evolution equations of the Z4 fields $\Te$ and $\ti{Z}^i$ take the form
\beq \label{eq:TeZsimp}
\dot{\Te} = -\, \al \[ K + \ka_1 \( 2 + \ka_2 \) \] \Te + \dots \, , \hspace{1cm} \dot{\ti{Z}}^i = -\, \al \( \frac{2}{3} \, K + \ka_1 \) \ti{Z}^i + \dots \, ,
\eeq
where the ellipses denote either second-order terms in perturbations around FLRW or source terms, so these are the ``linear" parts of the equations. We see that $\ka_1$ and $\ka_2$ are damping parameters, i.e. for appropriate values they push the system towards the constraint surface. For linear fluctuations around Minkowski space-time where $K=0$, demanding constraint stability leads to the bounds \cite{Gundlach:2005eh}
\beq
\ka_1 \geq 0 \, , \hspace{1cm} \ka_2 \geq - 1 \, .
\eeq
Around FLRW space-time, however, we have that $K < 0$ in our gauge, so the corresponding terms in \eqref{eq:TeZsimp} come with the wrong sign. Therefore, the ``undamped" CCZ4 system $\ka_1 = 0$ is not stable in the cosmological context, especially at early times where $|K|$ is large, because $\Te$ and $\ti{Z}^i$ diverge exponentially. Instead, the effectively undamped scheme in cosmology is the one corresponding to
\beq \label{eq:kappa}
\ka_1 = - \frac{2}{3}\, K \, , \hspace{1cm} \ka_2 = - \frac{1}{2} \, ,
\eeq
because this way the terms displayed in \eqref{eq:TeZsimp} cancel out. Greater values of $\ka_1$ would then reintroduce a damping effect. For the tests performed in this paper, we will exclusively work with \eqref{eq:kappa}. 

Finally, on the matter side we have a set of $N$ free-falling particles of mass $m$, with positions $x_a^i(t)$ and momenta $p_i^a(t)$, where $a = 1, \dots, N$. The derivation of the corresponding evolution equations is given in appendix \ref{app:partder}, the result being the geodesic equation in first-order form
\bea
\dot{x}_a^i & = & - \be^i + \al \chi E_a^{-1} \ti{\ga}^{ij} p_j^a \, , \label{eq:partevol} \\
\dot{p}^a_i & = & - E_a \pa_i \al + p_j^a \pa_i \be^j + \al E^{-1}_a \[ \chi \ti{\ga}^{jl} \ti{\Ga}^k_{il} p_j^a p^a_k - \frac{1}{2} \( E_a^2 - m^2 \) \chi^{-1} \pa_i \chi \] \, , \nn
\eea
where all gravitational fields appearing here are implicitly evaluated at $x^i_a$ and
\beq
E_a := \sqrt{m^2 + \chi \ti{\ga}^{ij}p_i^a p_j^a} \, ,
\eeq
is the energy of the $a$-th particle. The corresponding energy-momentum tensor components are also derived in appendix \ref{app:partder} and are given explicitly in their discretized version in the following subsection.

\subsection{Space discretization}

We discretize the field equations on a Cartesian mesh using finite difference methods. In particular, for the spatial derivatives we use a centered five-point stencil, i.e.
\beq
\pa_i f(\vec{x}) \to \frac{-2 f(\vec{x} + 2\vec{\De}_i) + 8f(\vec{x} + \vec{\De}_i) - 8f(\vec{x} - \vec{\De}_i) + 2 f(\vec{x} - 2\vec{\De}_i)}{12 \De x} \, , 
\eeq
where
\beq
(\vec{\De}_i)_j := \de_{ij} \De x \, ,
\eeq
with $\De x$ the lattice spacing. There are two exceptions to this. First, the $\pa_i$ appearing inside the convective derivative $\ced_t := \pa_t - \be^i \pa_i$ is replaced with the up/down-wind five-point stencil, depending on the sign of $\be^i$ 
\beq
\be^i \pa_i f(\vec{x}) \to \frac{\be^i}{12 \De x} \times \left\{ \begin{array}{ccc} f(\vec{x} + 3\vec{\De}_i) - 6 f(\vec{x} + 2\vec{\De}_i) + 18 f(\vec{x} + \vec{\De}_i) - 10 f(\vec{x}) - 3 f(\vec{x} - \vec{\De}_i) & {\rm if} & \be^i > 0 \\ - f(\vec{x} - 3 \vec{\De}_i) + 6 f(\vec{x} - 2 \vec{\De}_i) - 18 f(\vec{x} - \vec{\De}_i) + 10 f(\vec{x}) + 3 f(\vec{x} + \vec{\De}_i) & {\rm if} & \be^i < 0  \end{array} \right.  \, .
\eeq
Second, whenever we have double derivatives $\pa_i \pa_j$, the diagonal terms are replaced with the second derivative centered five-point stencil
\beq
\pa_i \pa_i f(\vec{x}) \to \frac{-f(\vec{x} + 2\vec{\De}_i) + 16f(\vec{x} + \vec{\De}_i) - 30f(\vec{x}) + 16f(\vec{x} - \vec{\De}_i) - f(\vec{x} - 2\vec{\De}_i)}{12 \De x^2} \, .
\eeq
As for the particle-mesh communication, the energy-momentum components are constructed by projecting the particle information according to 
\bea
E(\vec{x}) & := & \chi^{3/2}(\vec{x}) \sum_{a=1}^N E_a \we^{(3)} (\vec{x} - \vec{x}_a) \, , \nn \\
P_i(\vec{x}) & := & \chi^{3/2}(\vec{x}) \sum_{a=1}^N p_i^a \we^{(3)} (\vec{x} - \vec{x}_a) \, , \label{eq:EPS} \\
S_{ij}(\vec{x}) & := & \chi^{3/2}(\vec{x}) \sum_{a=1}^N \frac{p_i^a p_j^a}{E_a} \we^{(3)} (\vec{x} - \vec{x}_a) \, , \nn
\eea
where 
\beq
\we^{(3)}(\vec{x}) := \prod_{i = 1}^3 \we \, (x^i) \, , 
\eeq
and $\we$ denotes the triangle-shaped cloud function 
\beq
\we(x) := \left\{ \begin{array}{cc} \frac{3}{4} + \( \frac{x}{\De x} \)^2 & {\rm if} \,\,\, 0 \leq |x| \leq \frac{1}{2}\,\De x \\ \frac{1}{2} \( \frac{3}{2} - \frac{x}{\De x} \)^2 & {\rm if} \,\,\, \frac{1}{2}\, \De x \leq x \leq \frac{3}{2}\, \De x \\ 0 & {\rm otherwise} \end{array} \right. \, .
\eeq
For the interpolation of field values at particle positions we then use the inverse kernel.

\subsection{Time integration} \label{sec:timeint}

At the level of the FLRW space-time, our gauge corresponds to conformal time and comoving spatial coordinates, in terms of which light-like propagation corresponds to the same relation as in Minkowski space-time, i.e. $\De t = \De x$. For fluctuations around FLRW, the latter serves as a background space-time determining the causal structure of the dynamics, so it makes sense to consider a constant Courant factor in time
\beq
C := \frac{\De t}{\De x} < 1 \, ,
\eeq
which is the parameter relating the time step $\De t$ to the considered lattice spacing $\De x$. We have found that a satisfactory evolution is provided by a fourth-order Runge-Kutta (RK4) for the gravitational fields and a ``drift-kick-drift" for the particles. However, the particles are evolved only every $N_s$ cycles with time step $N_s \De t$. We found that constraint violation is significantly reduced when $N_s$ is around 10 and in this paper we will consider for definiteness $N_s = 8$.\footnote{Note that $N_s > 1$ poses no problem for the resolution of the particle dynamics, because their typical velocities are smaller than the speed of light by several orders of magnitude.} Given our particle integration method, the time step is therefore $N_s \De t$ for the particle positions and $2 N_s \De t$ for their momenta. Denoting by $G$ the set of gravitational fields $\{ \al, \pa_i \al, \be^i, \pa_i \be^i, \chi, \pa_i \chi, \ti{\ga}_{ij}, \ti{\Ga}^k_{ij} \}$ that must be interpolated at the particle positions \eqref{eq:partevol}, the time-integration loop is described as follows:

\begin{enumerate}

\item 
The particles are displaced according to (drift) 
\beq
\vec{x}_a(t + N_s\Delta t) = \vec{x}_a(t) + \dot{\vec{x}}_a \[ \vec{p}_a(t - N_s\De t), G(t,\vec{x}_a(t)) \] \times N_s\De t  \, .
\eeq

\item 
The matter fields $E$, $P_i$ and $S_{ij}$ are updated. 

\item 
The gravitational fields are evolved by $\De t$ through RK4 $N_s$ times. 

\item 
The particle momenta are updated according to (kick) 
\beq
\vec{p}_a(t + N_s\De t) = \vec{p}_a(t-N_s\De t) + \dot{\vec{p}}_a \[ \vec{p}_a(t - N_s\De t), G(t+N_s\De t, \vec{x}_a(t+N_s \De t)) \] \times 2N_s\De t \, . 
\eeq

\item 
The matter fields $E$, $P_i$ and $S_{ij}$ are updated. 

\item 
The gravitational fields are evolved by $\De t$ through RK4 $N_s$ times. 

\item 
The particles are displaced according to (drift)
\beq
\vec{x}_a(t + 2N_s\De t) = \vec{x}_a(t + N_s\De t) + \dot{\vec{x}}_a \[ \vec{p}_a(t + N_s\De t), G(t+2N_s\De t, \vec{x}_a(t+N_s \De t)) \] \times N_s\De t  \, . 
\eeq

\item
Send $t \to t + 2 N_s \De t$.

\end{enumerate}
\begin{figure}[htbp]
\begin{center}
\includegraphics[]{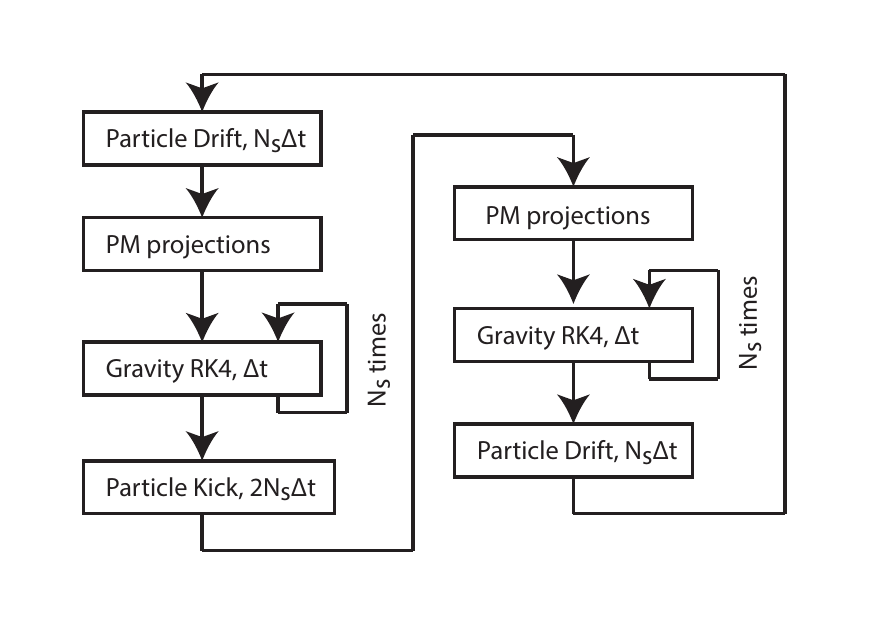} 
\caption{Time integration loop}
\label{fig:main_loop}
\end{center}
\end{figure}
An illustration of this loop is found in figure \ref{fig:main_loop}. Note that we impose the constraint \eqref{eq:Tconstraint} by hand
\beq
\ti{A}_{ij} \to \ti{A}_{ij} - \frac{1}{3}\, \ti{\ga}_{ij} \ti{\ga}^{kl} \ti{A}_{kl} \, ,
\eeq
at each sub-step of the RK4. Moreover, we include Kreiss-Oliger dissipation \cite{KO} for the evolution of the gravitational fields, choosing the sixth-order
one since we use a fourth-order time-integration. Thus, to the right-hand side of the evolution equation of some field $f$ we add
\bea
\De_{\rm KO} & := & \frac{\si}{64 \De x} \sum_{i=1}^3 \[ f(\vec{x} + 3\vec{\De}_i) - 6 f(\vec{x} + 2\vec{\De}_i) + 15 f(\vec{x} + \vec{\De}_i) - 20 f(\vec{x}) \right. \nn \\
 & & \hspace{3cm} \left. +\, 15 f(\vec{x} - \vec{\De}_i) - 6 f(\vec{x} - 2\vec{\De}_i) + f(\vec{x} - 3\vec{\De}_i)  \] \, .  
\eea
The normalization of the parameter $\si$ is such that the stability bounds are \cite{Alcubierre:1138167}
\beq
0 \leq \si \leq 2 C^{-1} \, .
\eeq
and here we will only consider the value $\si = 0.05$. Note that, when updating the matter fields \eqref{eq:EPS}, it is the particle-dependent part that is updated every $N_s$ steps, while the factor $\chi^{3/2}$ is updated at every step. Let us also point out the difference between this integration method and a staggered leapfrog where step 1 and 7 would be glued together. With our approach the computation of $x^i_a(t+2 N_s \De t)$ involves the gravitational fields evaluated at both $x^i_a(t)$ and $x^i_a(t + N_s\De t)$. This turns out to yield a better resolution of the particle dynamics and constraint violation control.

\section{Tests} \label{sec:tests}

\subsection{General definitions and specifications}

As far as the constraints are concerned, we will only display the Hamiltonian and momentum ones $H$ and $M_i$, respectively, given in \eqref{eq:H} and \eqref{eq:Mi}. We have monitored the rest of them $\{ D, \Te, \ti{Z}^i \}$\footnote{Remember that $D'$ is imposed algebraically as mentioned in the previous section.} as well and found that they are controlled better than $H$ and $M_i$. We will consider the absolute values $| H |$ and $| M | := \sqrt{\ga^{ij} M_i M_j}$, but also the more relevant relative quantities suggested in \cite{Mertens:2015ttp} 
\beq \label{eq:HrMr}
H_r := \frac{|H|}{\sqrt{\sum_n T^H_n T^H_n}} \, , \hspace{1cm}  M_r := \frac{|M|}{\sqrt{\sum_n \ga^{ij} T^M_{n,i} T^M_{n,j}}} \, ,
\eeq
where $T^H_n$ and $T^M_{n,i}$ denote the $n$-th term appearing on the right-hand side of \eqref{eq:H} and \eqref{eq:Mi}, respectively. These relative errors therefore capture the number of significant digits at which the cancellation in $H$ and $M_i$ occurs. Given some error field $\ph(\vec{x})$ on the lattice, the measures we will output are 
\beq \label{eq:Lmeas}
L_{\infty}[\ph] := \max_g |\ph(\vec{x}_g)| \, , \hspace{1cm} L_1[\ph] := \frac{1}{N_g^3} \sum_g |\ph(\vec{x}_g)| \, ,
\eeq
where $g$ indexes the grid points and $N_g^3 := (L/\De x)^3$ is their total number. In all three tests there is some relation to the FLRW solution which we consider in the case of zero spatial curvature and zero cosmological constant. On this solution the non-zero field components are 
\beq \label{eq:FLRW}
\al = a \, , \hspace{1cm} \chi = a^{-2} \, , \hspace{1cm} K = - 3 a^{-1} \cH \, , \hspace{1cm} \ti{\ga}_{ij} = \de_{ij} \, , \hspace{1cm} E = 3 a^{-2} \cH^2 \, , 
\eeq
where
\beq
\cH := \frac{\dot{a}}{a} \, ,
\eeq
is the conformal Hubble parameter and we have chosen the normalization $Q = 1$ (see \eqref{eq:alsol}). Remember that, given our choice of lapse \eqref{eq:BMal}, $t$ is conformal time.\footnote{In cosmological perturbation theory the latter is usually denoted by ``$\ta$" or ``$\et$" and its derivative is denoted by a prime, but here this is the variable with respect to which we solve our equations numerically, which is why we use $t$ and a dot, respectively. Indeed, from the perspective of numerical relativity, it makes no sense to change the symbol for the time variable only because we have specified the lapse function.} For all simulations our initial time will always be $t = 0$ and 
\beq
a(t_f) = 1 \, , \hspace{1cm} \cH(t_f) = H_0 \, ,
\eeq
where $t_f$ is the final time. With these we have 
\beq \label{eq:aH}
a(t) = \[ \frac{t + \sqrt{a(0)} \( t_f - t \)}{t_f} \]^2 \, ,  \hspace{1cm} \cH(t) = \frac{H_0}{\sqrt{a(t)}} \, ,
\eeq
and we will express time-evolution either with respect to $t$, or the corresponding FLRW redshift
\beq
z := \frac{1}{a} - 1 \, .
\eeq
In our units ${\rm Mpc}/h = c = 1$ the Hubble constant is 
\beq
H_0 := 100 \( h/{\rm Mpc} \) \( {\rm km}/{\rm s} \) \equiv 100 \( {\rm km}/{\rm s} \) \approx 3.336\times 10^{-4} \, .
\eeq
All of our runs start at $z(0) = 1000$, so the corresponding final time $(z = 0)$ is
\beq
t_f = \frac{2}{H_0} \( 1 - \frac{1}{\sqrt{1 + z(0)}} \) \approx 5800 \, .
\eeq
Note, however, that $H_0$ and the redshift parametrization cannot be given their realistic interpretations, because we are considering a pure-matter universe. Nevertheless, the corresponding cosmology has the correct orders of magnitude and we have access to strong inhomogeneity by going up to $z = 0$. 

Finally, we will provide no details on how we generate initial particle data $x_a^i(0)$ and $p_i^a(0)$ that reproduce the desired fields $E(0)$ and $P_i(0)$, as this will be addressed in another paper of this series. Let us just say that we consider regularly distributed particles with respect to the lattice, before displacing them to obtain the initial positions. The corresponding mass is then determined by
\beq \label{eq:mass}
m = 3 H_0^2\, \frac{\De x^3}{N_{\rm ppc}} \approx 3.3 \times 10^{-7} \frac{\De x^3}{N_{\rm ppc}} \, ,
\eeq
where $N_{\rm ppc}$ denotes the number of particles per grid cell before displacement. The number $N_{\rm ppc}$ will always be considered constant, meaning that the total number of particles scales as $N \sim \De x^{-3}$.

\subsection{FLRW robustness test}

\begin{figure}[htbp]
\includegraphics[width=0.495\columnwidth]{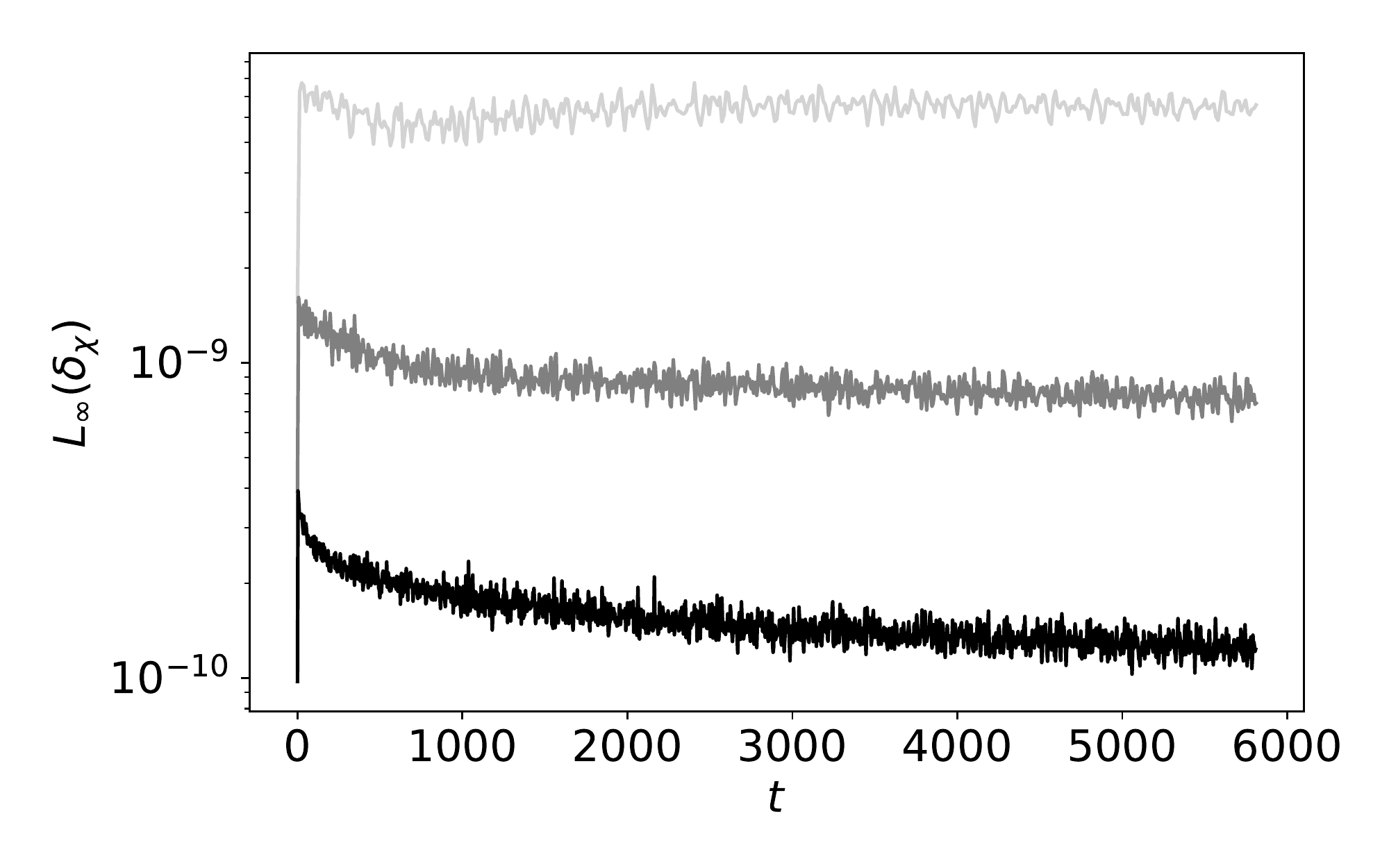} 
\includegraphics[width=0.495\columnwidth]{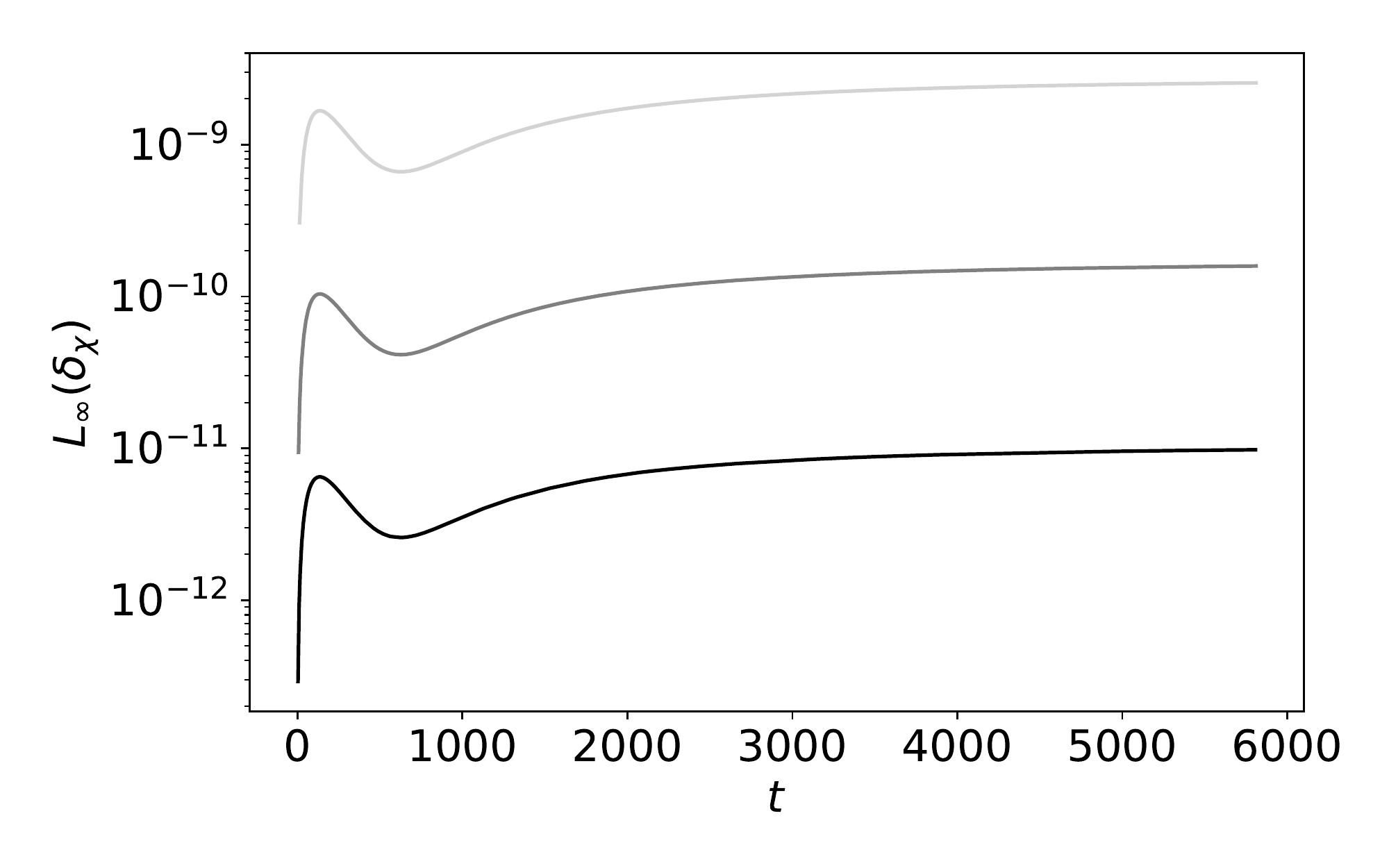} 
\caption{The $L_{\infty}$ measure of the relative error $\de_{\chi}$ for the FLRW solution with noise (left) and without (right) at three resolutions $\De x = 8$ (light gray), $\De x = 4$ (gray) and $\De x = 2$ (black).}
\label{fig:flrw_chi}
\end{figure}

\begin{figure}[htbp]
\includegraphics[width=0.495\columnwidth]{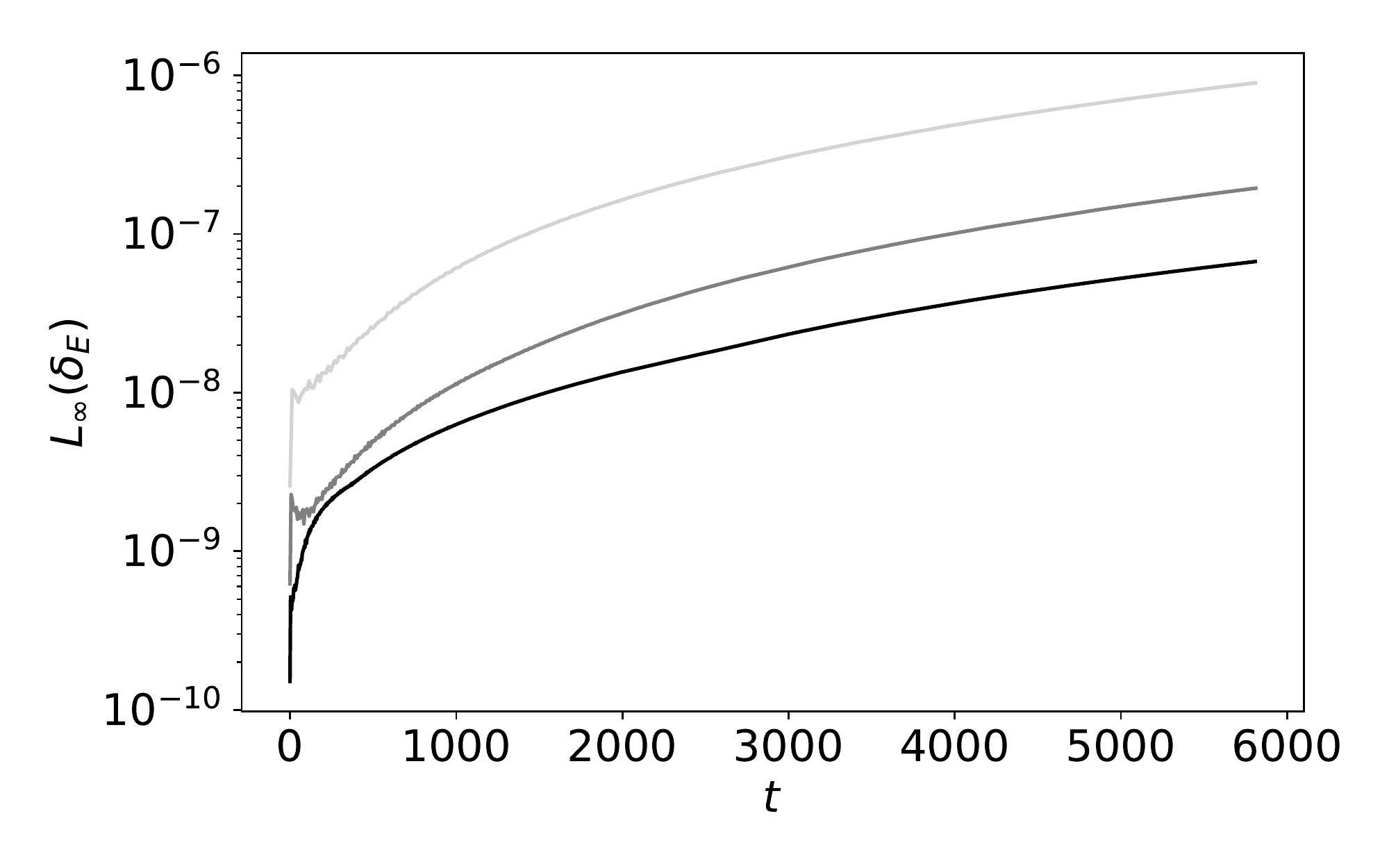} 
\includegraphics[width=0.495\columnwidth]{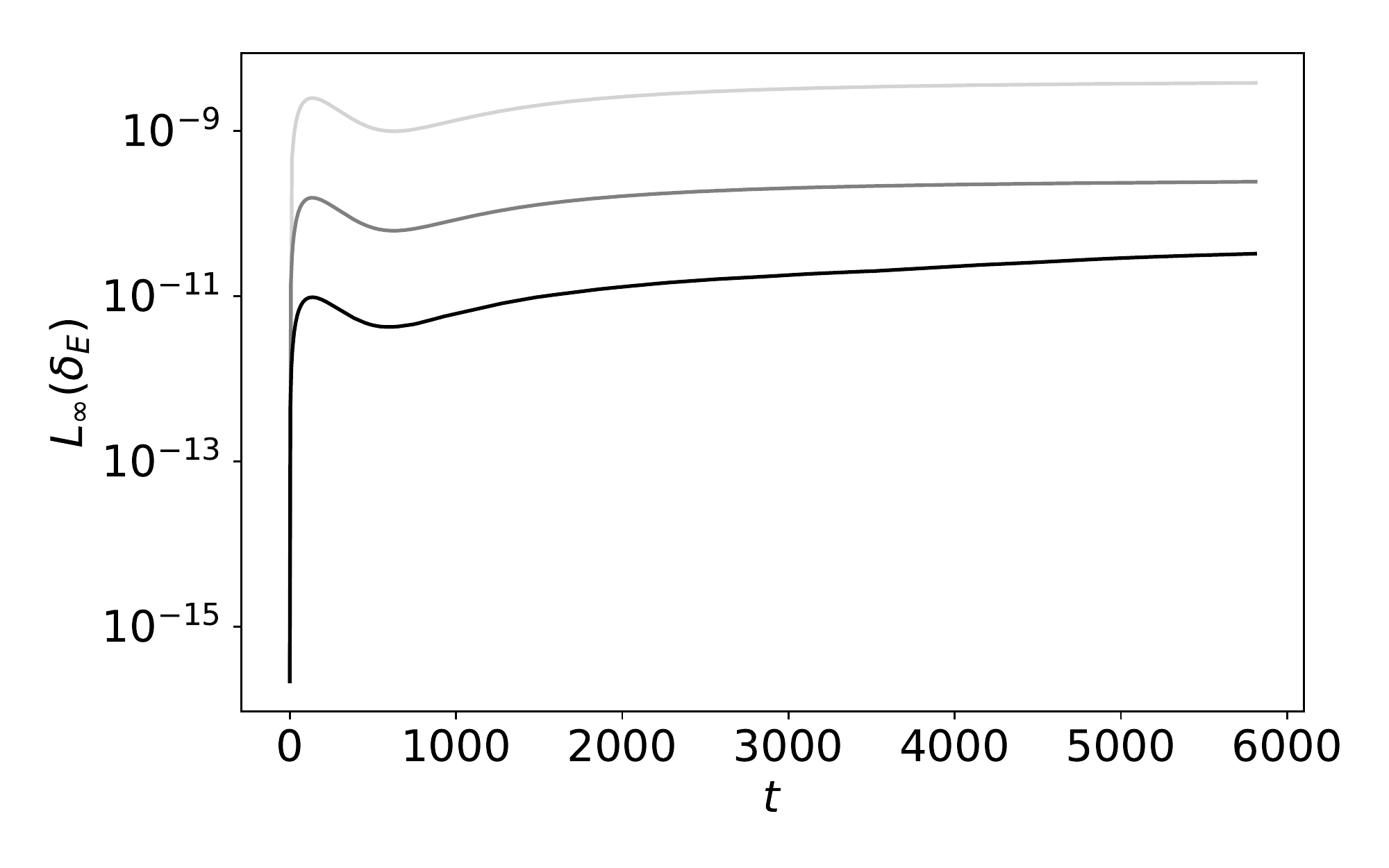} \\
\center{\includegraphics[width=0.495\columnwidth]{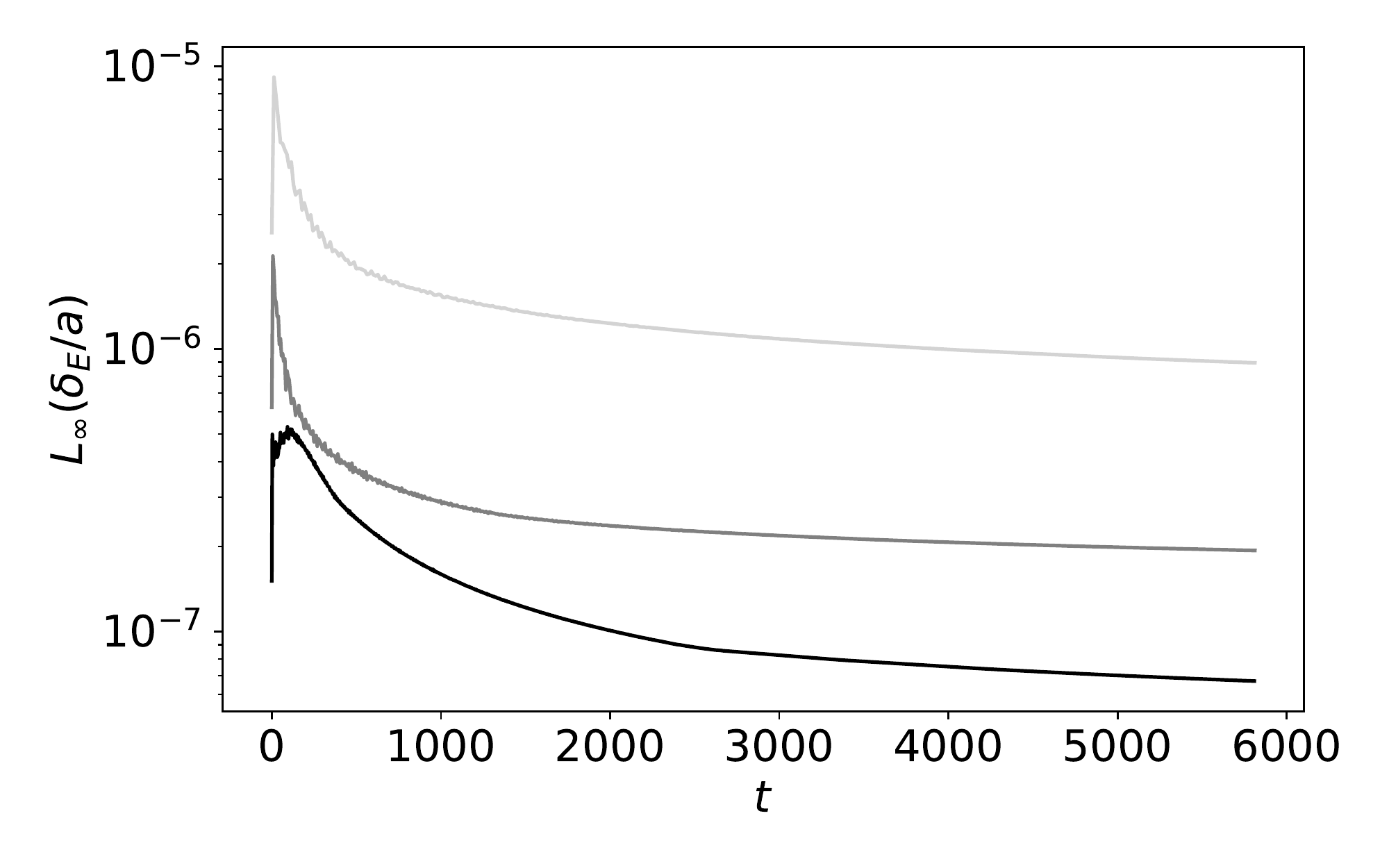}}
\caption{The $L_{\infty}$ measure of the relative error $\de_E$ for the FLRW solution with noise (top-left) and without (top-right) at three resolutions $\De x = 8$ (light gray), $\De x = 4$ (gray) and $\De x = 2$ (black). At the bottom we plot $\de_E/a$ for the former case to show that the noise grows slower than what is expected from linear perturbation theory, i.e. a $\sim a$ behavior.}
\label{fig:flrw_E}
\end{figure}

\begin{figure}[htbp]
\includegraphics[width=0.495\columnwidth]{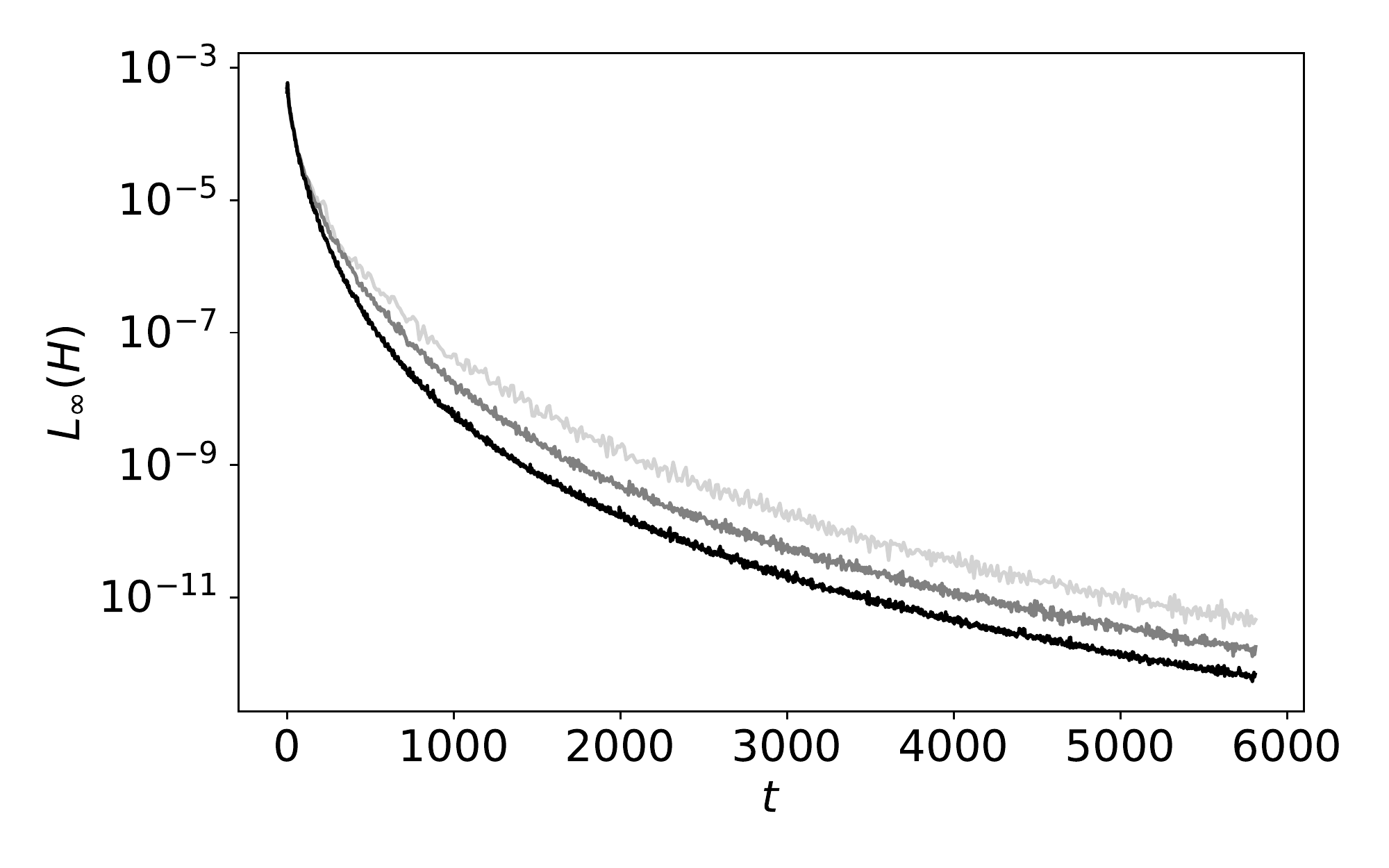} 
\includegraphics[width=0.495\columnwidth]{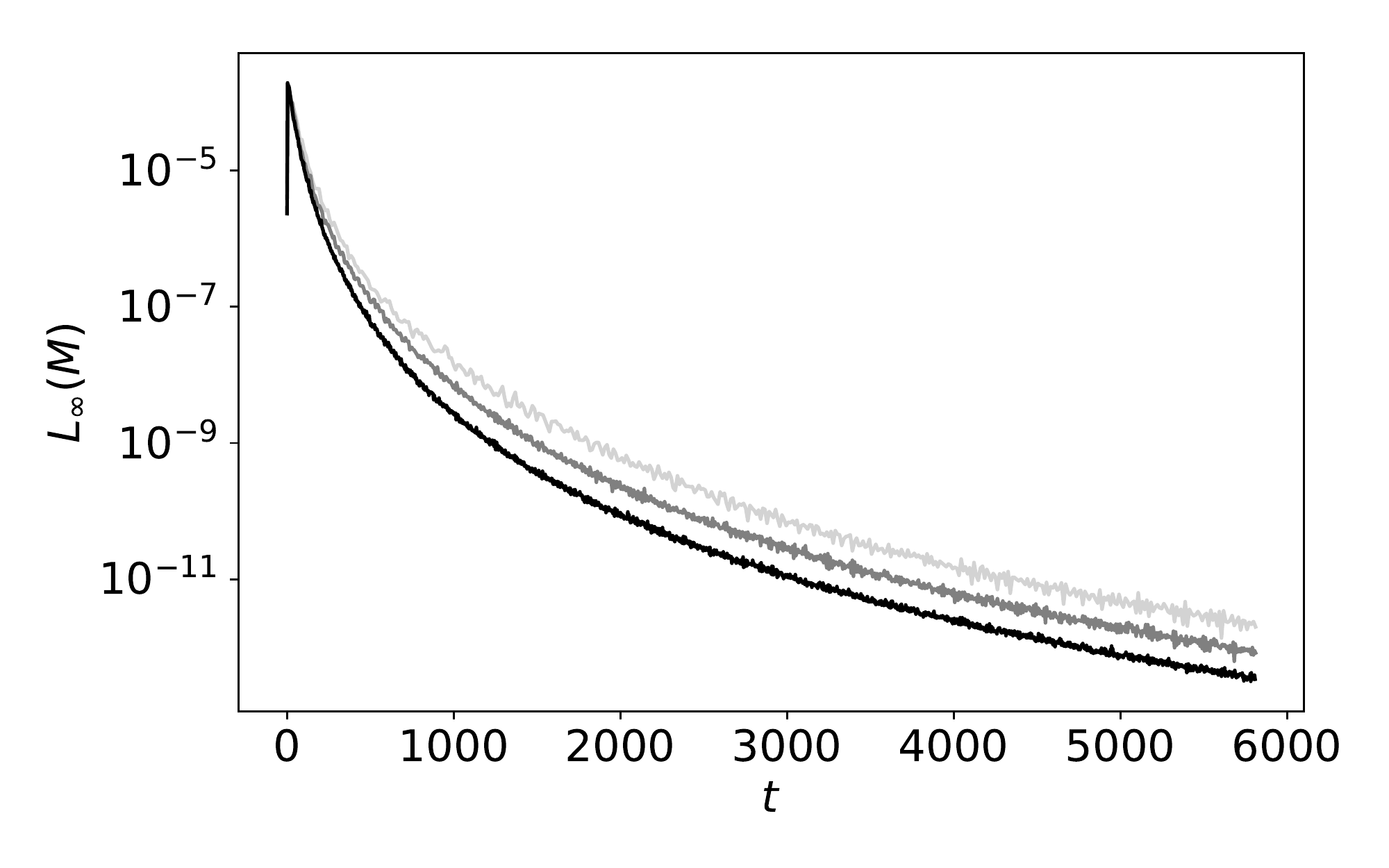} 
\caption{The $L_{\infty}$ measure of the Hamiltonian $H$ (left) and momentum $M$ (right) constrains for the FLRW solution with noise at three resolutions $\De x = 8$ (light gray), $\De x = 4$ (gray) and $\De x = 2$ (black).}
\label{fig:flrw_constraints}
\end{figure}

Here we conduct a robustness test, as defined in \cite{Alcubierre:2003pc, Babiuc:2007vr}, but adapted to the FLRW solution instead of Minkowski and also to the inclusion of particles. We thus consider the following perturbation of the FLRW initial conditions
\bea 
\al(0,\vec{x}) & = & a(0) \[ 1 + \ep(\vec{x}) \] \, , \\
\chi(0,\vec{x}) & = & a^{-2}(0) \[ 1 + \ep(\vec{x}) \] \, , \\
\Te(0,\vec{x}) & = & \ep(\vec{x}) \, , \\
K(0,\vec{x}) & = & -3 (a^{-1} \cH)(0) \[ 1 + \ep(\vec{x}) \] \, , \\
\hat{\Ga}^i(0,\vec{x}) & = & \ep^i(\vec{x}) \, , \\
\ti{\ga}_{ij}(0,\vec{x}) & = & \de_{ij} + \ep_{ij}(\vec{x}) \, , \\
\ti{A}_{ij}(0,\vec{x}) & = & \ep_{ij}(\vec{x}) \, ,
\eea
while for the particles we have
\beq
\de x^i_a(0) = \ep_a^i \, ,\hspace{1cm} p_i^a(0) = m \ep_i^a \, ,
\eeq
where $\de x_a^i$ is the displacement from the regular configuration. The $\ep_{\dots}$ numbers are drawn randomly out of a uniform distribution independently for each component, for each point $\vec{x}$ for the fields and for each particle. The amplitude of the distribution is $10^{-7} N_g^{-2}$ for the fields and $10^{-7} N_g^{-3}$ for the particles.\footnote{For the fields the $\sim N_g^{-2}$ dependence is required because of the presence of second-order spatial derivatives, which grow like $\sim N_g^2$ on random noise. For the particles, we have that the initial displacement field $\de x^i(0)$ is related to the corresponding density contrast through $\pa^2 \de x^i \sim \pa_i \de$. Thus, since we keep $N_{\rm ppc}$ fixed, the amplitude of $\de$ grows with resolution increase like $\sim N_g$, so the $\de x^i$ amplitude must scale as $\sim N_g^{-3}$ for $\de$ to follow the same trend as the other fields $\sim N^{-2}_g$. Indeed, we have checked that there is no convergence for the error on $E$ if the particle amplitude follows $\sim N_g^{-2}$ instead of $\sim N_g^{-3}$.} The runs are performed in a box with comoving size $L = 64$ at three spatial resolutions $\De x \in \{ 8,4,2 \}$, meaning that $N_g \in \{ 8, 16, 32 \}$. The Courant factor is $C = 0.1$ and we use one particle per grid cell $N_{\rm ppc} = 1$. Figures \ref{fig:flrw_chi} and \ref{fig:flrw_E} show the relative errors of $\chi$ and $E$ with respect to their respective analytical solutions
\beq
\de_{\chi} := \left| \frac{\chi - \chi_{\rm FLRW}}{\chi_{\rm FLRW}} \right| \, , \hspace{1cm} \de_E := \left| \frac{E - E_{\rm FLRW}}{E_{\rm FLRW}} \right| \, ,
\eeq
while figure \ref{fig:flrw_constraints} shows the absolute constraint violations $H$ and $M$. In all cases we plot the $L_{\infty}$ measure. We see that $\de_E$ grows in the presence of noise, which can be understood by the fact that this quantity has a growing mode $\sim a$ already at the analytical level, i.e. any inhomogeneity must grow. In the bottom panel of figure \ref{fig:flrw_E}, we plot the evolution of $\de_E/a$ and see that it is bounded in time, so that the noise is under control in this particular context. The overall verdict is that we are able to follow the analytical solution with good stability and convergence.

\subsection{Scalar linear mode test}

In this test we check whether the code can accurately evolve a single scalar mode of inhomogeneity in the linear regime around the FLRW solution. We use the definitions, residual gauge choices and the Zel'dovich condition described in appendix \ref{app:perts}. The initial conditions are therefore completely determined by the gauge-invariant density contrast and here we consider a single mode profile
\beq
\de_{\star}(0,x,y,z) = A \sin \frac{2 \pi x}{L} \, ,
\eeq
where $L$ is the comoving box size. We choose $A = 10^{-7}$, which will lead to $\de_{\star} \sim 10^{-4}$ at redshift zero, thus remaining inside the regime of validity of linear perturbation theory at all times. Moreover, we choose $L = 128$, meaning that the mode starts outside the initial comoving Hubble radius $\cH^{-1}(0) \approx 95$ and finishes inside. The rest of the parameters are $C = 0.1$, $N_{\rm ppc} = 27$ and $\De x \in \{ 4, 2, 1 \}$. 

\begin{figure}[htbp]
\includegraphics[width=0.495\columnwidth]{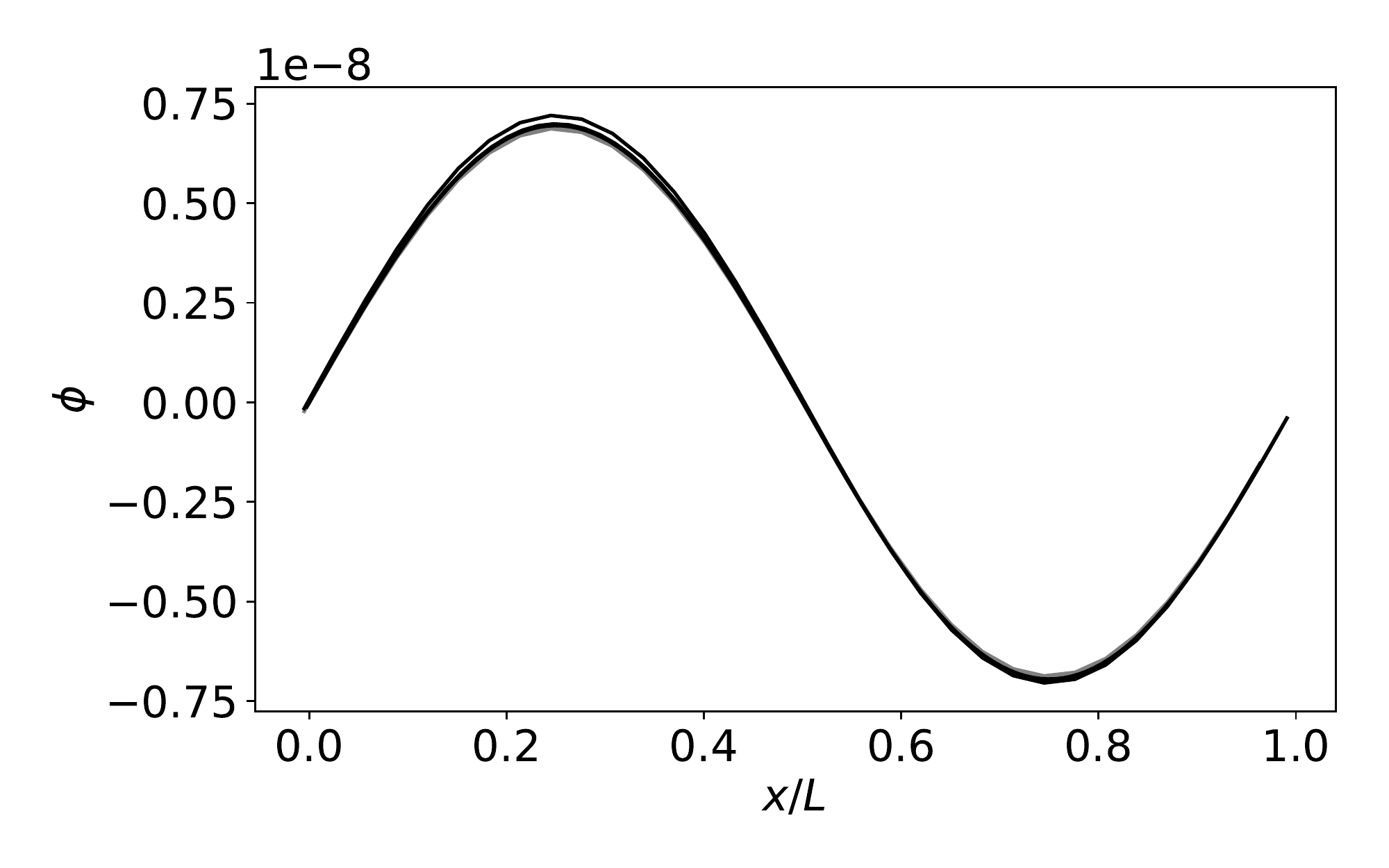} 
\includegraphics[width=0.495\columnwidth]{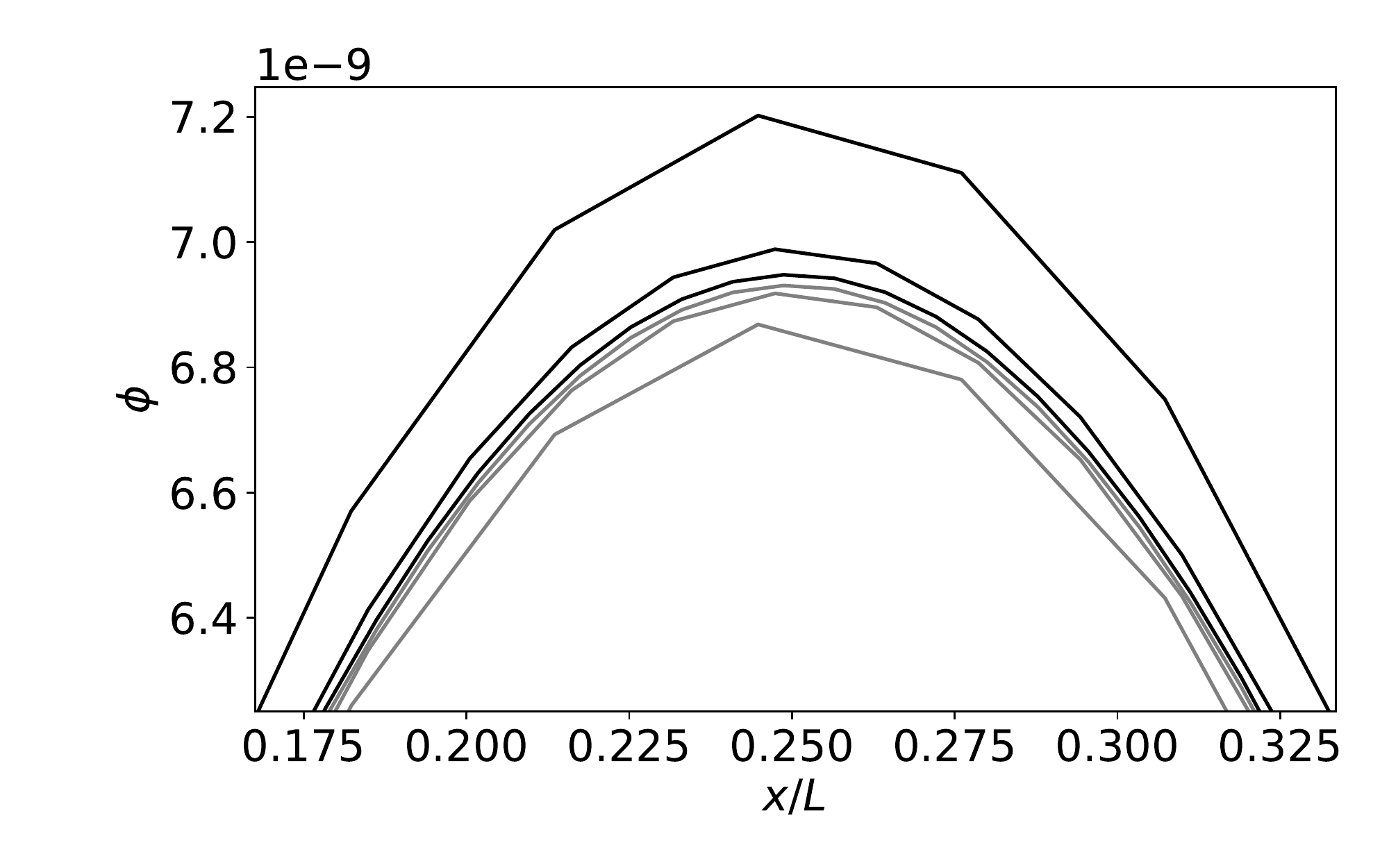} 
\caption{The gravitational potential $\vph$ profile (left) for the scalar linear mode solution and a zoom of the non-trivial region (right). The plot contains both the initial (gray) and final (black) profiles at all three resolutions.}
\label{fig:sm_phi}
\end{figure}

\begin{figure}[htbp]
\includegraphics[width=0.495\columnwidth]{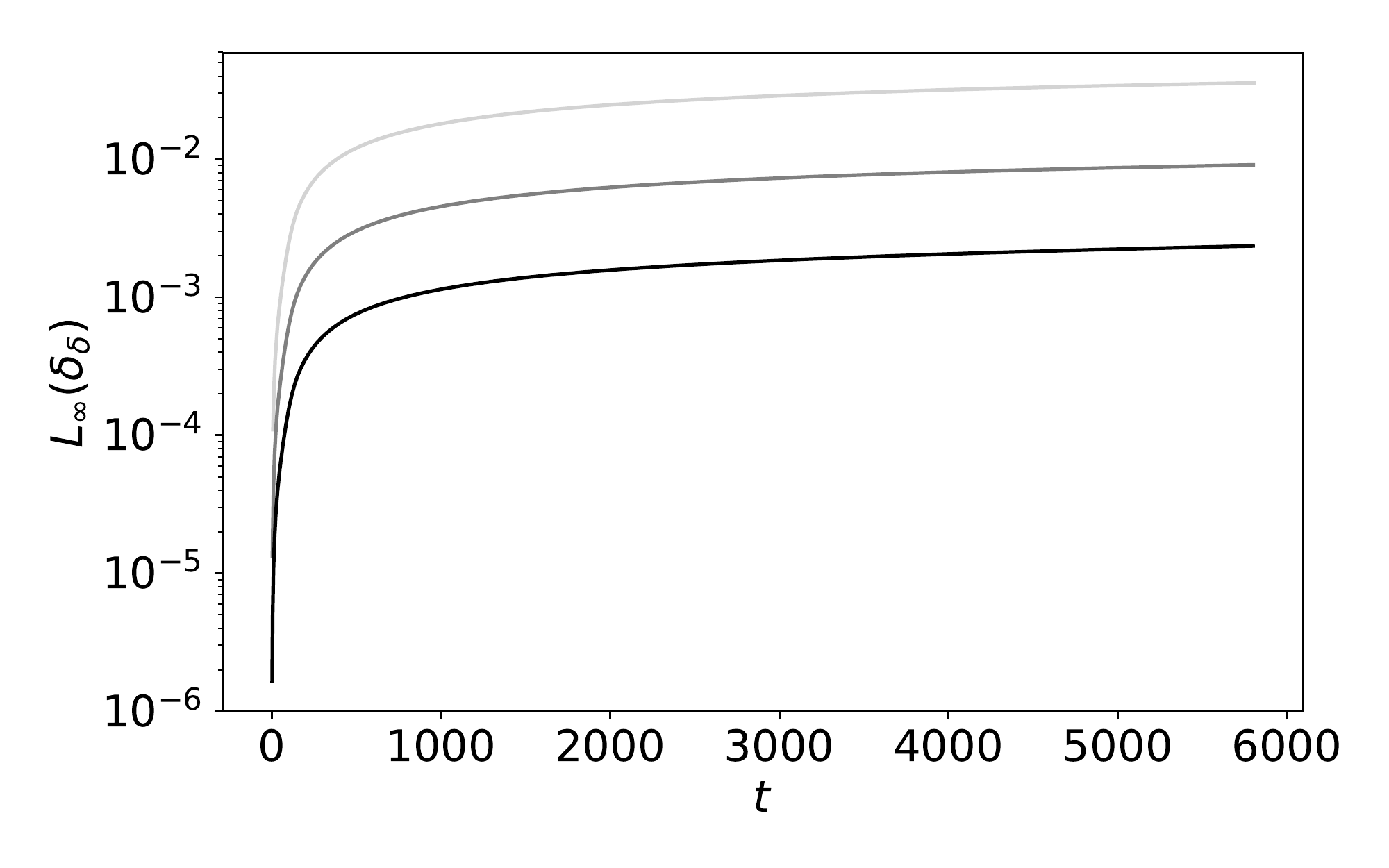}
\includegraphics[width=0.495\columnwidth]{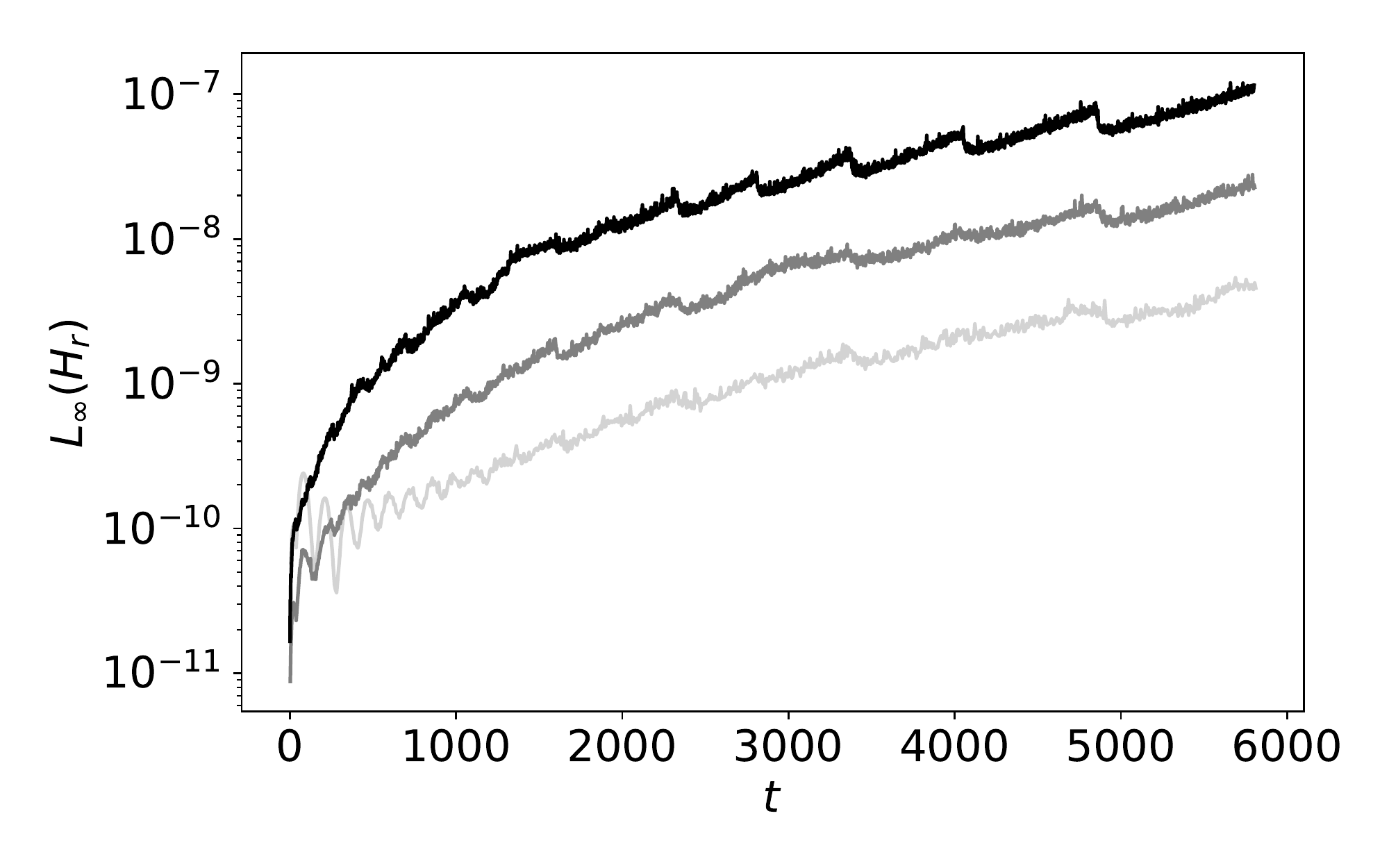} 
\caption{The $L_{\infty}$ measures of the relative error of the energy density contrast $\de_{\de}$ (left) and the relative constraint violation $H_r$ (right) for the scalar linear mode solution at three resolutions $\De x = 4$ (light gray), $\De x = 2$ (gray) and $\De x = 1$ (black).}
\label{fig:sm_de_Hr}
\end{figure}

Note that the gravitational potential $\vph$ must remain constant in time, which is what we see in figure \ref{fig:sm_phi}, where we have plotted its profile for all three resolutions at both the initial and final times. The right panel is a magnification of the region around the maximum and shows that the initial and final profiles converge towards each other with increasing resolution, although from opposite sides. On the left panel of figure \ref{fig:sm_de_Hr} we plot the relative error of $\de$ with respect to the amplitude of the analytical solution 
\beq
\de_{\de} := \frac{|\de_{\rm num.} - \de_{\rm an.}|}{L_{\infty} \( \de_{\rm an.} \)} \, ,
\eeq
which is controlled and converges with resolution. On the right panel we plot the $L_{\infty}$ measure of the relative Hamiltonian constraint $H_r$. We see that it diverges with resolution and, in fact, this is the case for all the constraints, for both the absolute and relative cases and for both the $L_{\infty}$ and $L_1$ measures. Note that this divergence under resolution increase also occurs in the linearized (gravitational) wave test around Minkowski space-time \cite{Daverio:2018tjf} for the BSSNOK, CCZ4, Z4cc and Z4c schemes (with and without constraint violation damping) for three different gauge choices (see \cite{Daverio:2018tjf} for details). However, this is not observed in the linear regime of the typical cosmology test (multi-mode) of the next subsection, which is the relevant one for cosmology. It therefore seems that the present divergence is an artefact of the plane symmetry of this special configuration.\footnote{This divergence is also not observed in the triple mode tests performed in \cite{Pretorius:2018lfb, East:2019chx, Giblin:2018ndw}, which supports this explanation of the problem. However, there might be also other factors involved in this issue, because in \cite{Giblin:2018ndw} the authors also perform planar-symmetric single mode tests and obtain convergence. They use a BSSNOK scheme, but with different amplitude, wave-length and redshift range, and also a more sophisticated particle-mesh communication method.} Moreover, despite this divergence of $L_{\infty}(H_r)$, its magnitude is still several orders of magnitude smaller that the relative error on $\de$.

\subsection{Typical cosmology test}

In this subsection we test the behavior of our code for initial conditions that exhibit the typical inhomogeneities one encounters in cosmology. As in the previous test, we consider again the scalar linear perturbation theory around the FLRW solution with the residual gauge choices and the Zel'dovich condition described in appendix \ref{app:perts}, but now only for our initial conditions at redshift 1000. For the initial gauge-invariant density contrast $\de_{\star}$, we use the power spectrum provided by the linear Boltzmann code {\tt CLASS} \cite{Lesgourgues:2011re, Blas:2011rf} to generate a corresponding random field, which is then used to determine the initial field and particle data.\footnote{We have checked that our initial conditions respect the Zel'dovich approximation well enough $\vph + h \sim 10^{-3} \vph$ and also that the divergence of the momentum dominates its curl by $|\vec{k} \times \vec{P}| \sim 10^{-3} \vec{k} \cdot \vec{P}$ on the resolved scales.} 

We will consider two simulations: one with box size $L = 704$ and one with $L = 256$, cutting-off the power at wavelengths $L_{\rm cut} = 220$ and $L_{\rm cut} = 40$, respectively, and with resolutions $\De x \in \{ \frac{L}{32}, \frac{L}{64}, \frac{L}{128} \}$ for the former and $\De x \in \{ \frac{L}{64}, \frac{L}{128}, \frac{L}{256} \}$ for the latter. Thus, the cut-off scales $L_{\rm cut}$ correspond to ten times the lattice spacing of the poorest resolution. Note that these cut-offs are imposed at the initial conditions, but evolution will generate structure at smaller scales. The figures \ref{fig:z100} and \ref{fig:z0} correspond to the smaller box simulation $L = 256$.  The rest of the parameters are a Courant factor of $C = 0.05$ and twenty-seven particles per grid cell $N_{\rm ppc} = 27$. We will also denote by 
\beq
C_{\De x}[X] := \log_2 \frac{X_{2\De x}}{X_{\De x}} \, ,
\eeq
the convergence ratio of a given quantity $X$, where $X_{\De x}$ denotes the value computed with resolution $\De x$. Since we have three resolutions for each run, we will have two $C_{\De x}[X]$ values for each quantity $X$. Note that, although the field derivatives are computed with fourth-order precision, the particle time integration is of second order, so a successful convergence corresponds to $C_{\De x}[X] \geq 2$. 

\begin{figure}[htbp]
\includegraphics[width=0.495\columnwidth]{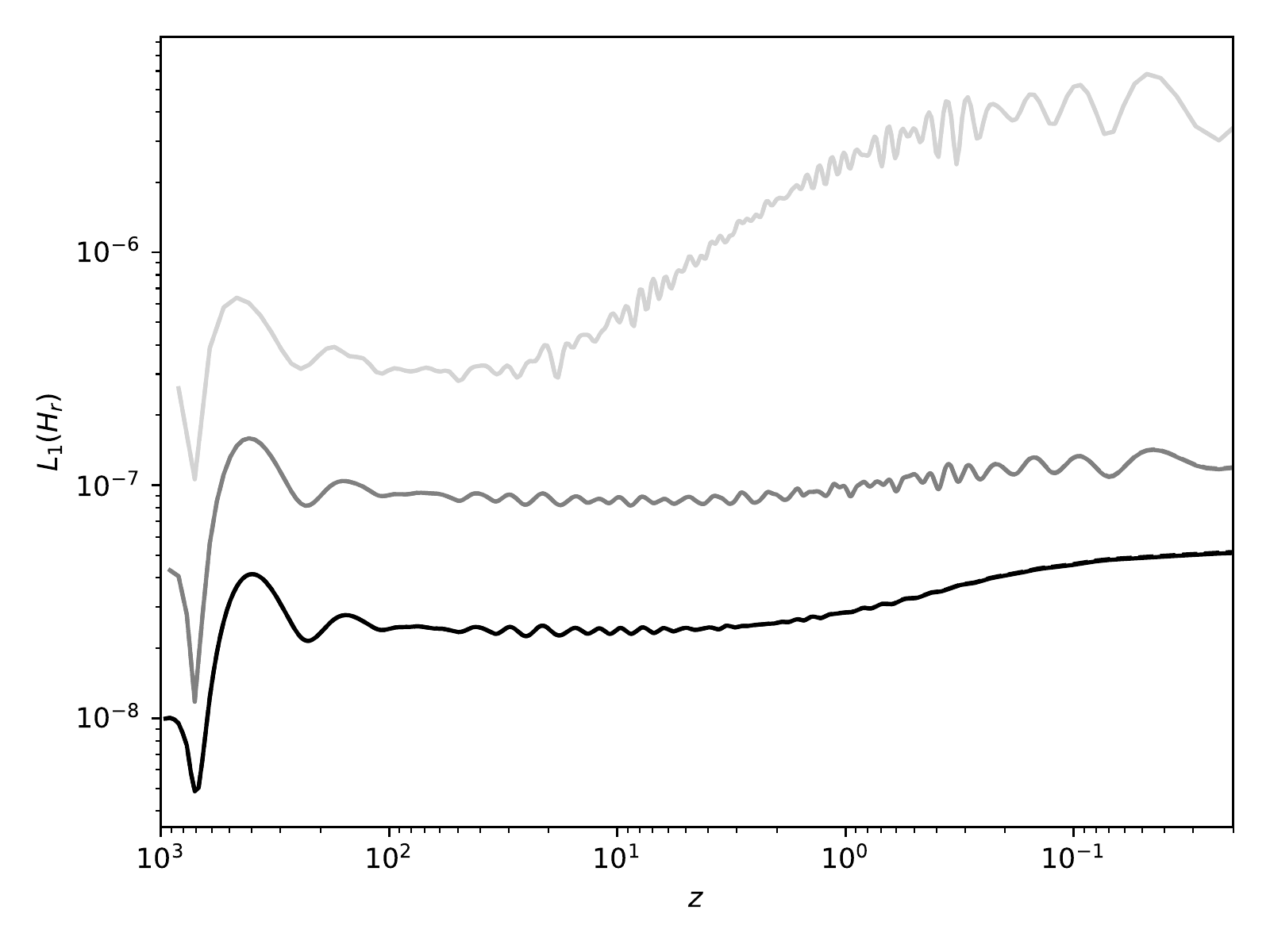} 
\includegraphics[width=0.495\columnwidth]{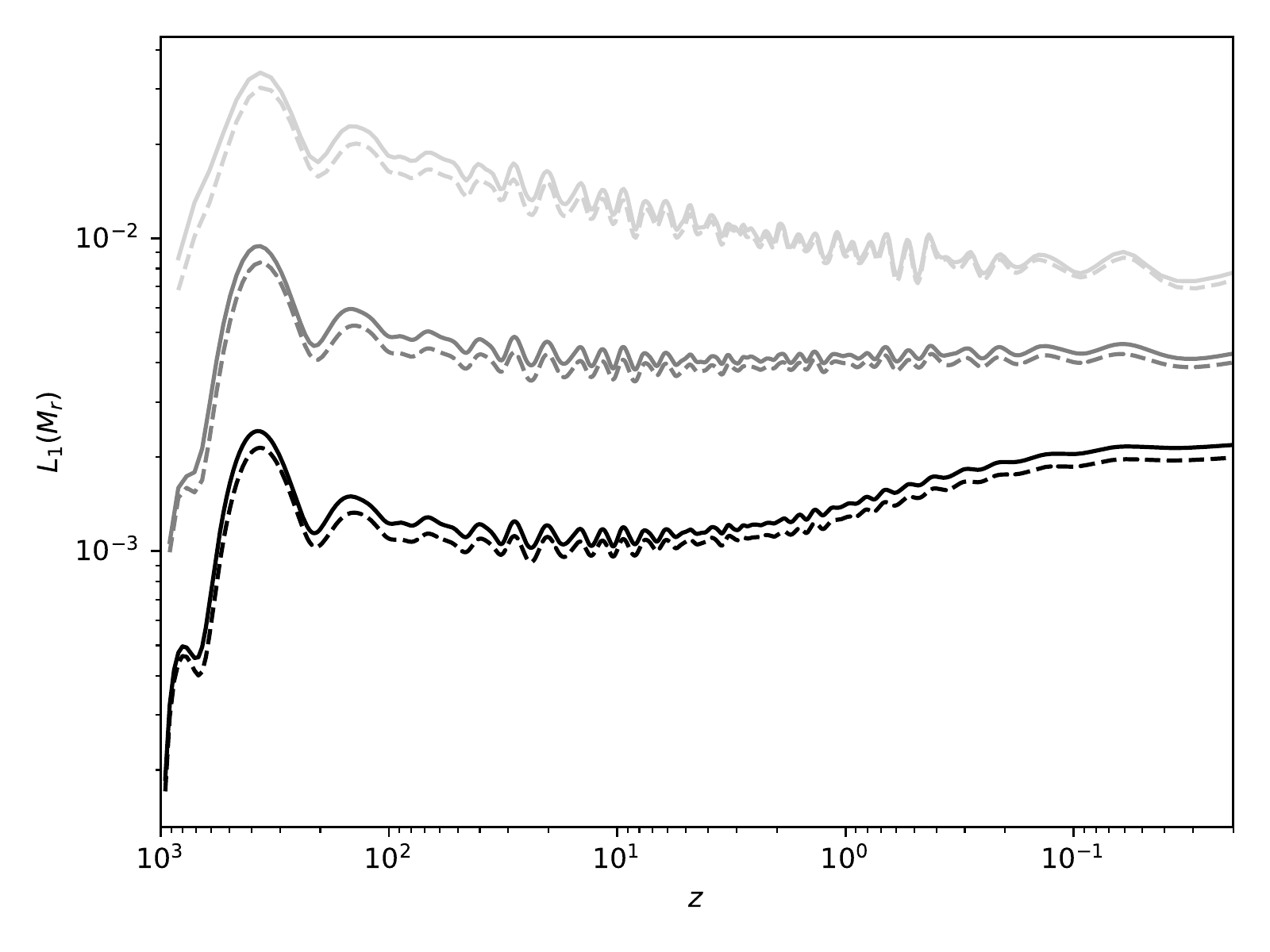} 
\caption{The $L_1$ measure of the relative Hamiltonian constraint $H_r$ (left) and momentum constraint $M_r$ (right) for the typical cosmology test with $L=704$ as a function of redshift for three resolutions: $\De x = L/32$ (light gray), $\De x = L/64$ (gray) and $\De x = L/128$ (black). The dashed lines correspond to the ratio of averages measure used in \cite{Macpherson:2018btl}. For the Hamiltonian constraint the two measures are indistinguishable.}
\label{fig:cosmo_mac_constr}
\end{figure}

\begin{figure}[htbp]
\includegraphics[width=0.495\columnwidth]{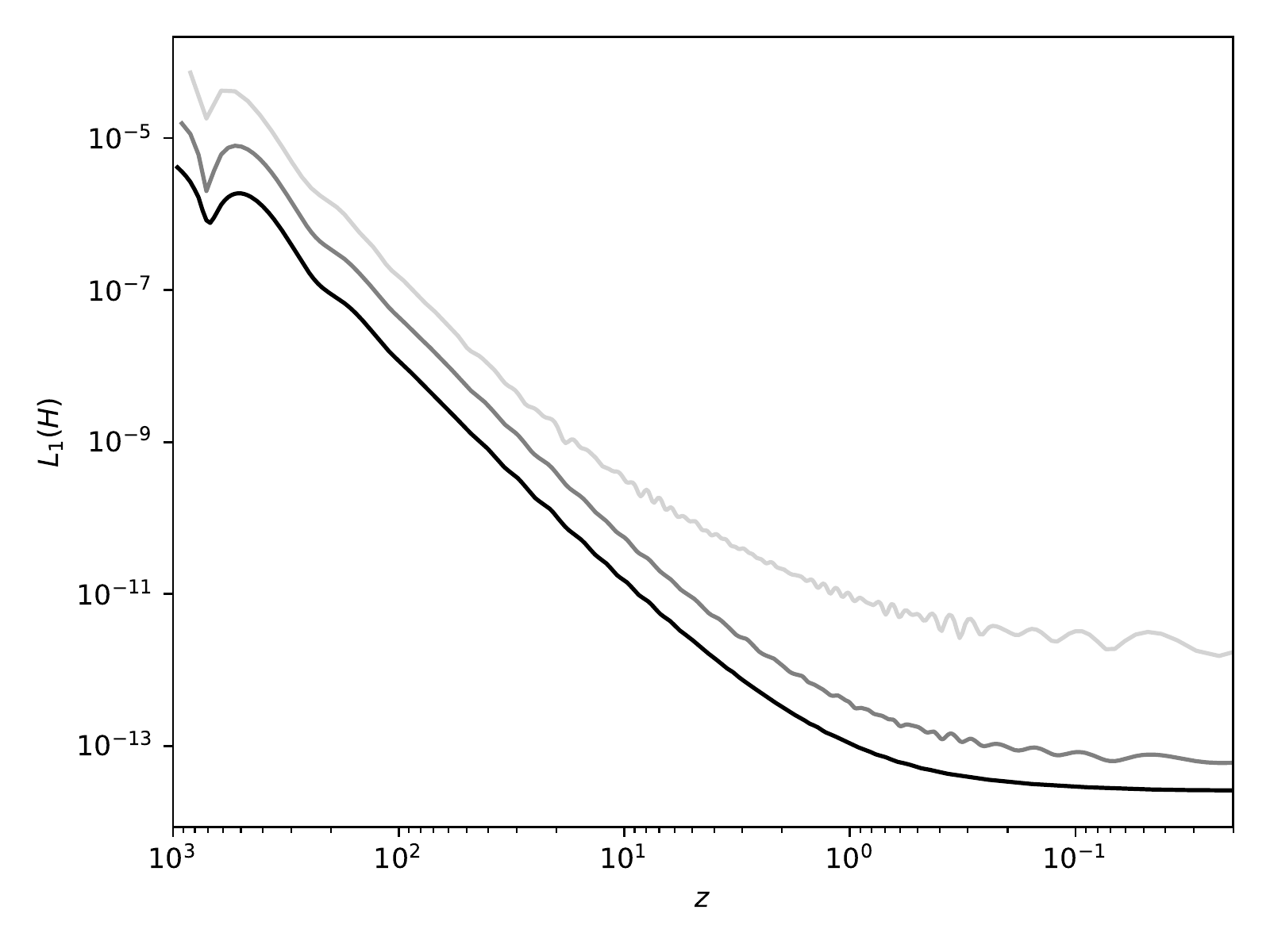} 
\includegraphics[width=0.495\columnwidth]{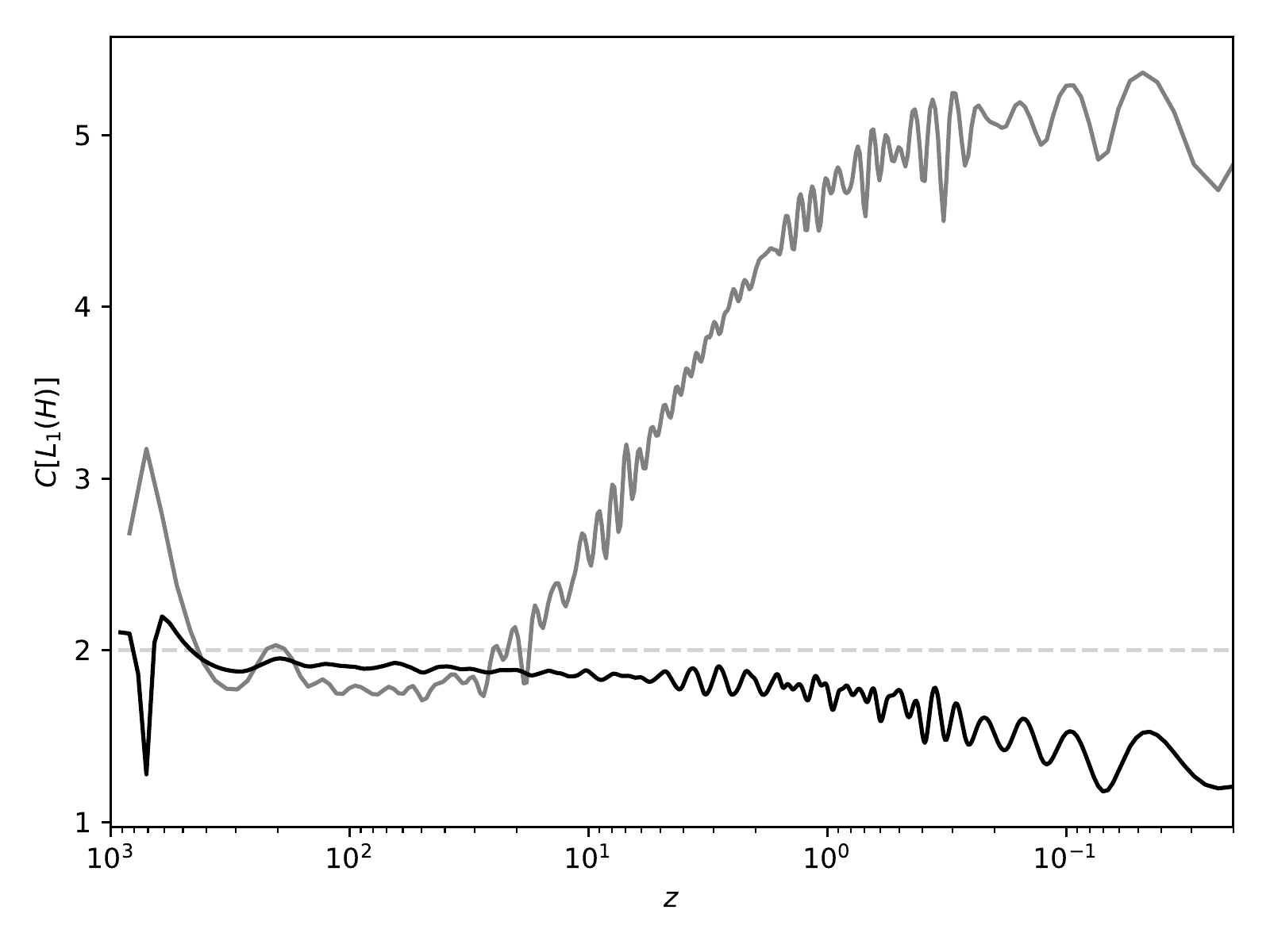} 
\caption{The $L_1$ measure of the absolute Hamiltonian constraint $H$ (left) and its converge ratio $C[L_1(H)]$ (right) as a function of redshift for the typical cosmology test with $L=704$. The left plot displays the three resolutions: $\De x = L/32$ (light gray), $\De x = L/64$ (gray) and $\De x = L/128$ (black), while the right plot displays the two resolution ratios $C_{L/64}[L_1(H)]$ (gray) and $C_{L/128}[L_1(H)]$ (black).}
\label{fig:cosmo_mac_H}
\end{figure}

\begin{figure}[htbp]
\includegraphics[width=0.495\columnwidth]{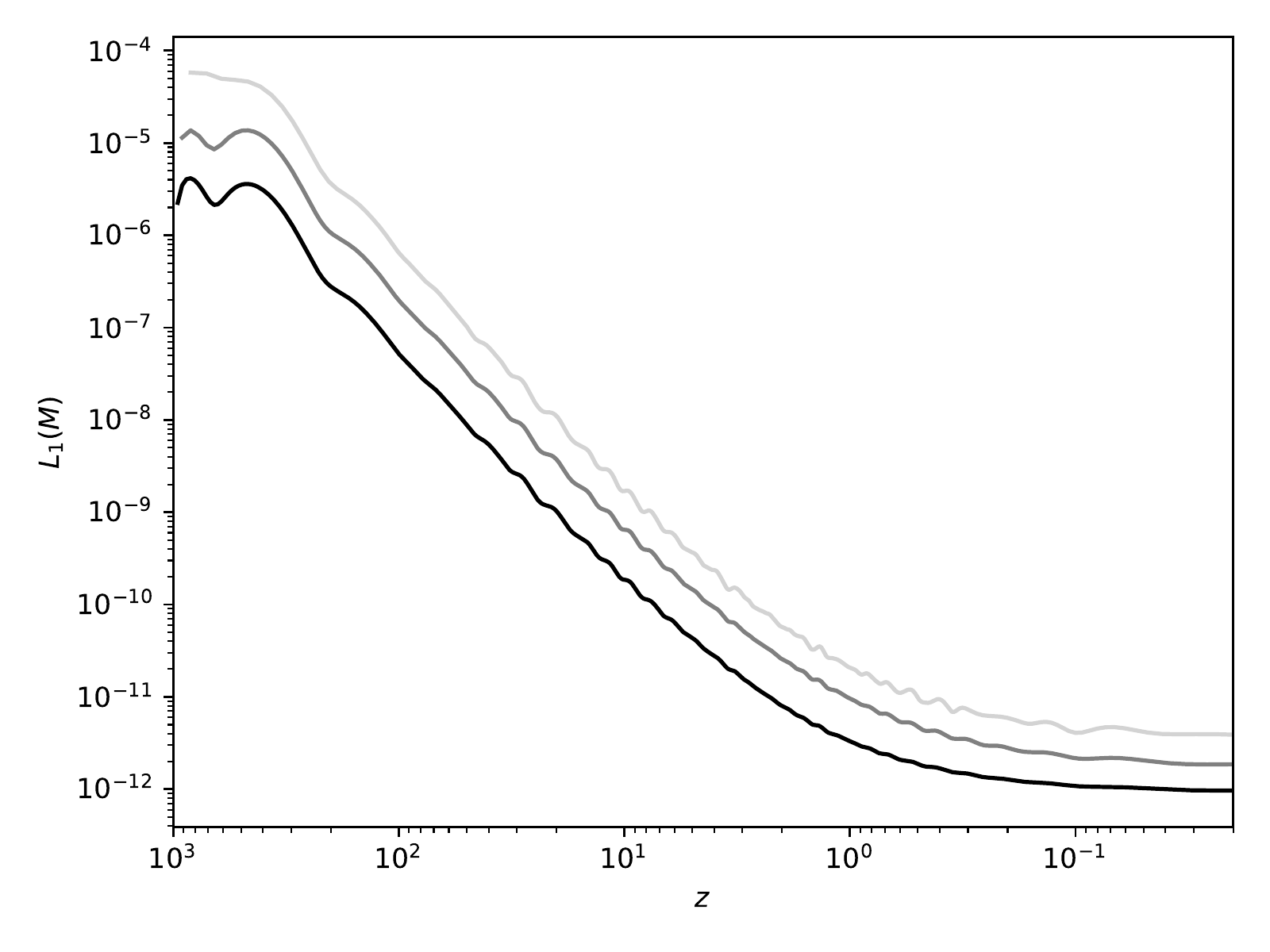} 
\includegraphics[width=0.495\columnwidth]{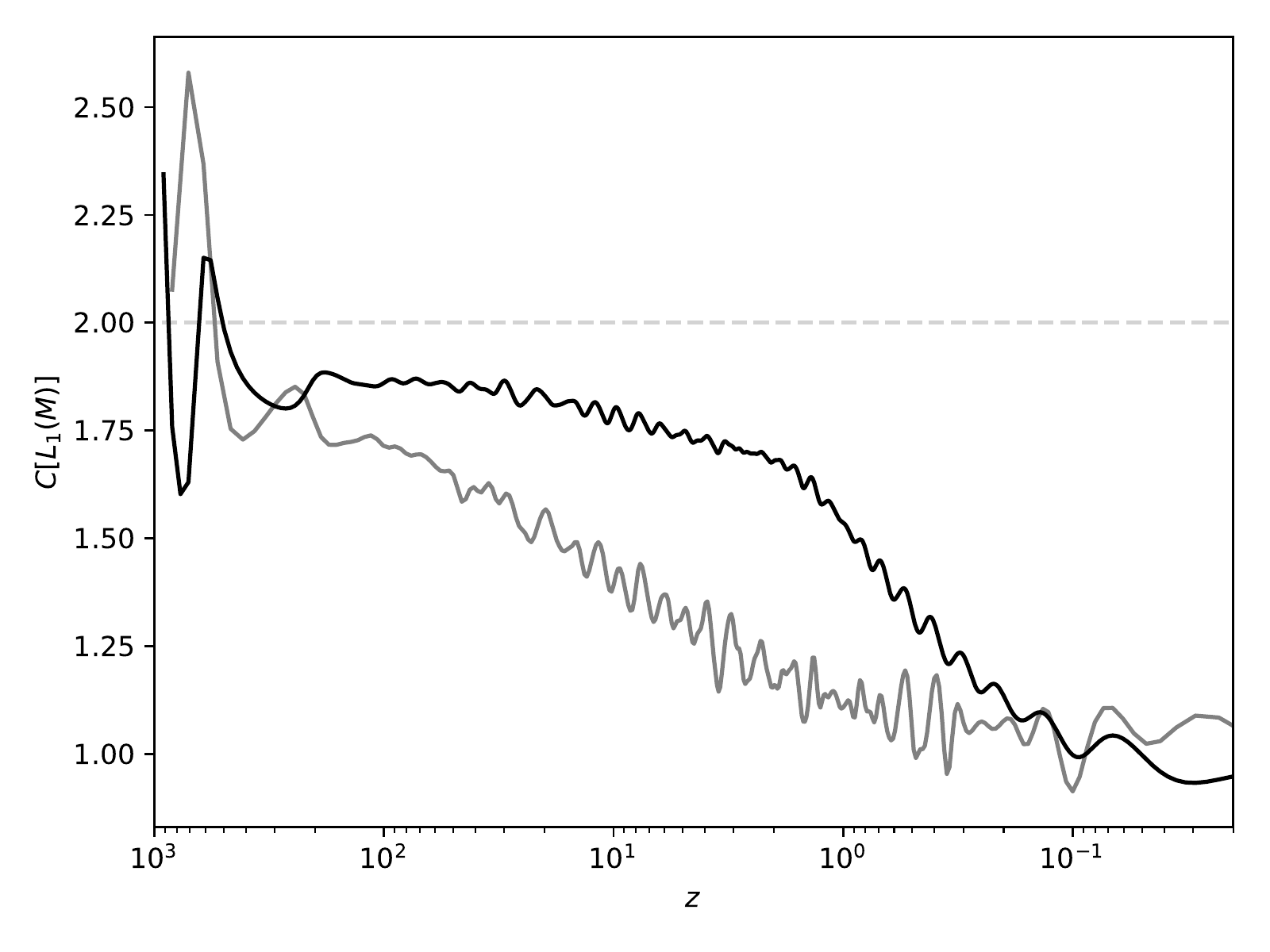} 
\caption{The $L_1$ measure of the absolute momentum constraint $M$ (left) and its converge ratio $C[L_1(M)]$ (right) as a function of redshift for the typical cosmology test with $L=704$. The left plot displays the three resolutions: $\De x = L/32$ (light gray), $\De x = L/64$ (gray) and $\De x = L/128$ (black), while the right plot displays the two resolution ratios $C_{L/64}[L_1(M)]$ (gray) and $C_{L/128}[L_1(M)]$ (black).}
\label{fig:cosmo_mac_M}
\end{figure}

In the $L = 704$ case we are mainly considering scales at which shell-crossing is negligible, so the particle dynamics should coincide with the ones of a pressureless perfect fluid. We have chosen the particular numbers $L = 704$ and $L_{\rm cut} = 220$ in order to compare with the perfect fluid NR simulation performed in \cite{Macpherson:2018btl}.\footnote{Note that in \cite{Macpherson:2018btl} the units are ${\rm Mpc} = 1$ instead of Mpc$/h = 1$, as we have here, and the authors use $h = 0.704$.} More precisely, this matches the simulations of \cite{Macpherson:2018btl} with ``controlled number of modes" for which the constraints are plotted as a function of redshift. Moreover, we work with the same gauge as \cite{Macpherson:2018btl}. In figure \ref{fig:cosmo_mac_constr} we plot the $L_1$ measure of the relative Hamiltonian and momentum constraints \eqref{eq:HrMr} as a function of redshift. Note that the $L_1$ measure of \cite{Macpherson:2018btl}, given in their equation (C4), is the ratio of averages, instead of the average of ratios which we use \eqref{eq:Lmeas}, and these two do not obey a definite order relation. Nevertheless, we verify that they are of the same order of magnitude by displaying both. In figures \ref{fig:cosmo_mac_H} and \ref{fig:cosmo_mac_M} we plot the $L_1$ measure of the absolute constraints $H$ and $M$, respectively, along with the corresponding convergence ratios.

Let us start with the magnitude of constraint violation, thus focusing on figure \ref{fig:cosmo_mac_constr} and the right panels of figure 11 of \cite{Macpherson:2018btl}. We first note that our relative constraints are quite stable in time (except for $H_r$ of the poorest resolution run), as opposed to the relative Hamiltonian constraint of \cite{Macpherson:2018btl} which grows until it reaches a plateau value. Moreover, we have three orders of magnitude less error for $L_1(H_r)$ and one order of magnitude less error for $L_1(M_r)$ at redshift zero. As for the convergence of the constraint violation, we compare our figures \ref{fig:cosmo_mac_H} and \ref{fig:cosmo_mac_M} to the left panels of figure 11 and to figure 12 of \cite{Macpherson:2018btl}. We find that our Hamiltonian constraint behaves less well, in that it is not really converging at second order after $z \sim 1$. For the momentum, we obtain a clearer separation of the curves at all times, but end up with a convergence of first order only. Our verdict is therefore that we are able to control constraint violation a lot better than in \cite{Macpherson:2018btl} and that our convergence over the full evolution is of comparable quality. It must be stressed, however, that \cite{Macpherson:2018btl} employ the BSSNOK scheme without constraint-damping mechanisms.

\begin{figure}[htbp]
\includegraphics[width=0.495\columnwidth]{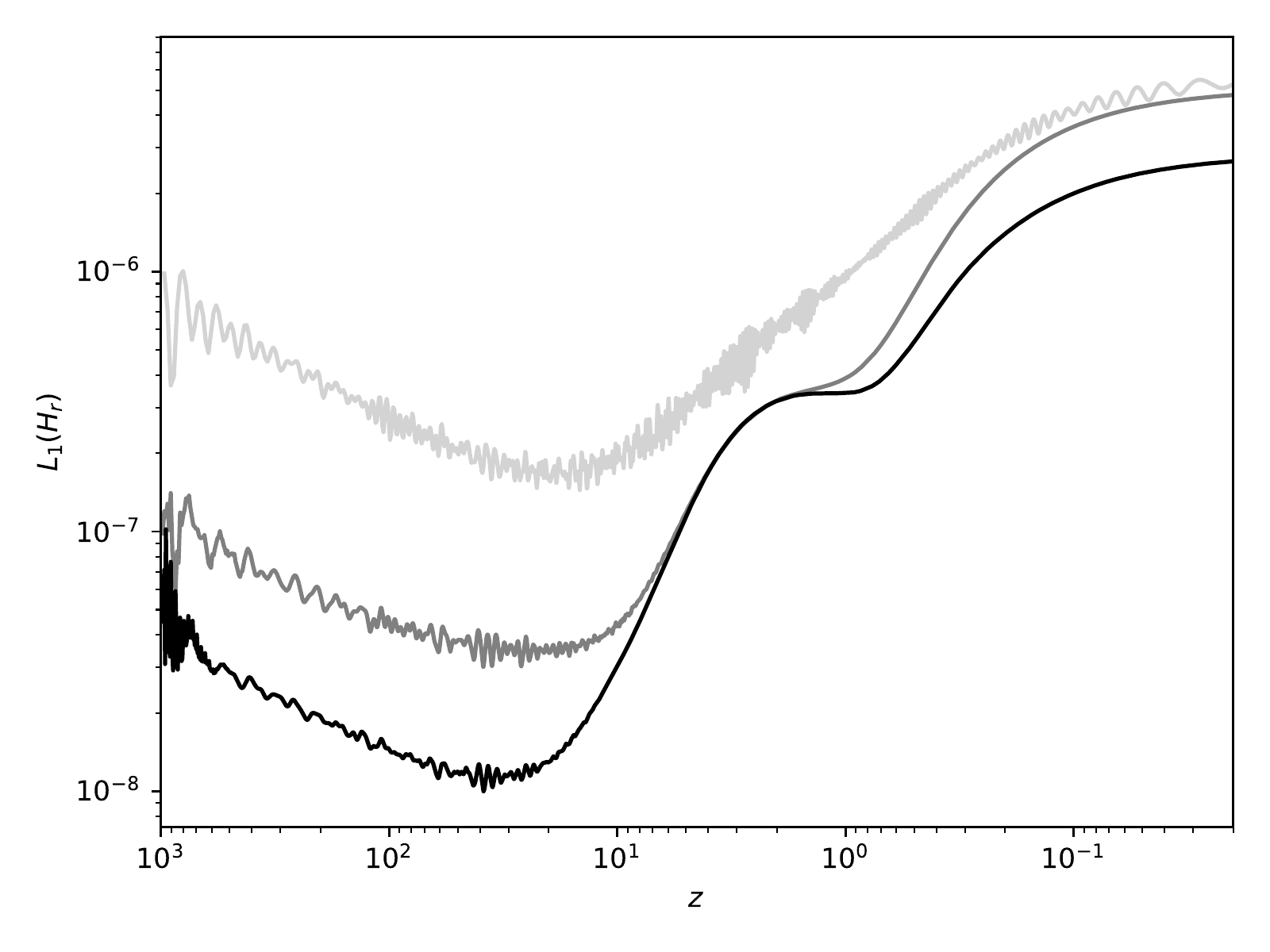} 
\includegraphics[width=0.495\columnwidth]{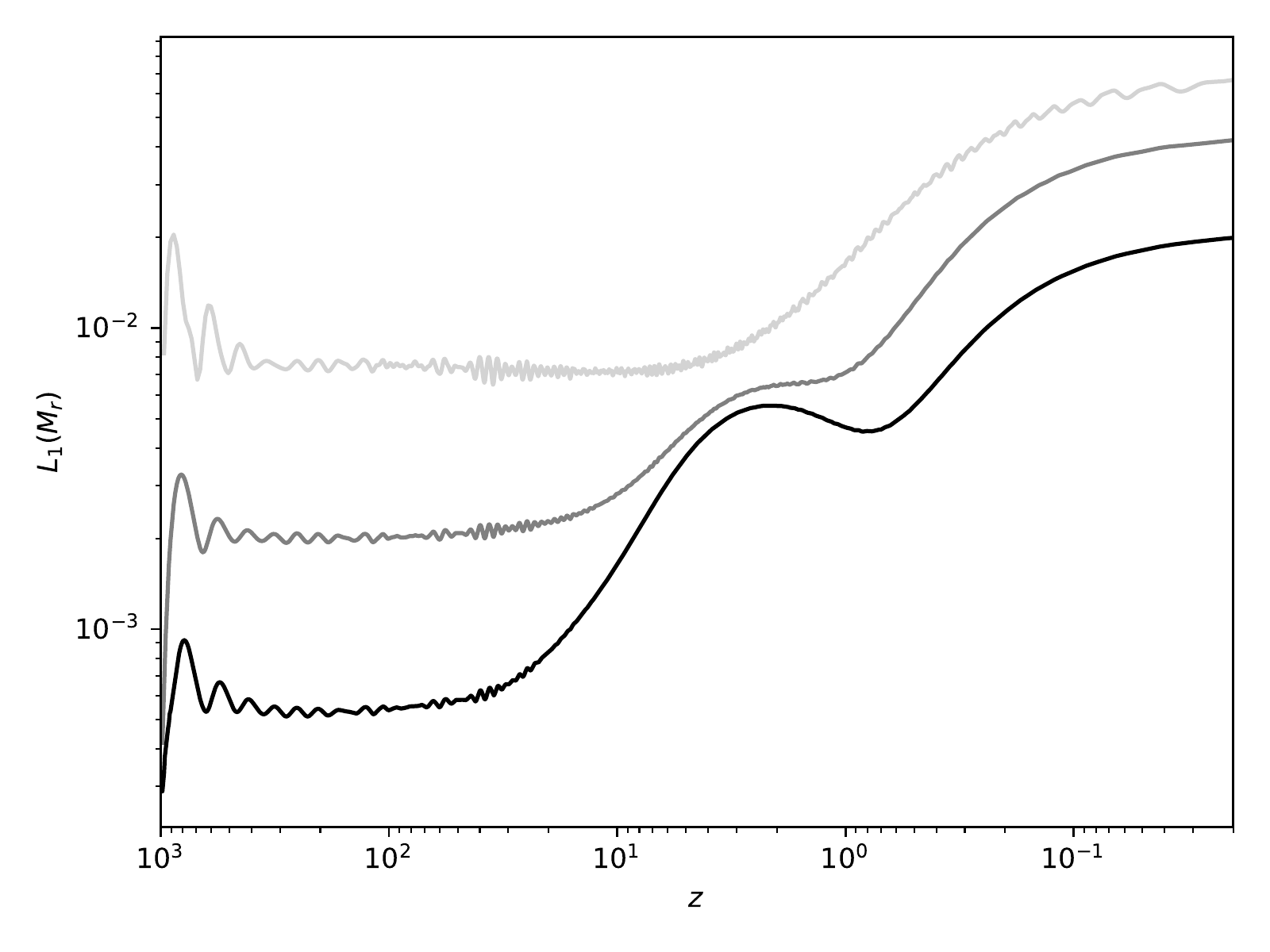} 
\caption{The $L_1$ measure of the relative Hamiltonian constraint $H_r$ (left) and momentum constraint $M_r$ (right) for the typical cosmology test with $L=256$ as a function of redshift for three resolutions: $\De x = L/64$ (light gray), $\De x = L/128$ (gray) and $\De x = L/256$ (black). }
\label{fig:cosmo_constr}
\end{figure}

\begin{figure}[htbp]
\includegraphics[width=0.495\columnwidth]{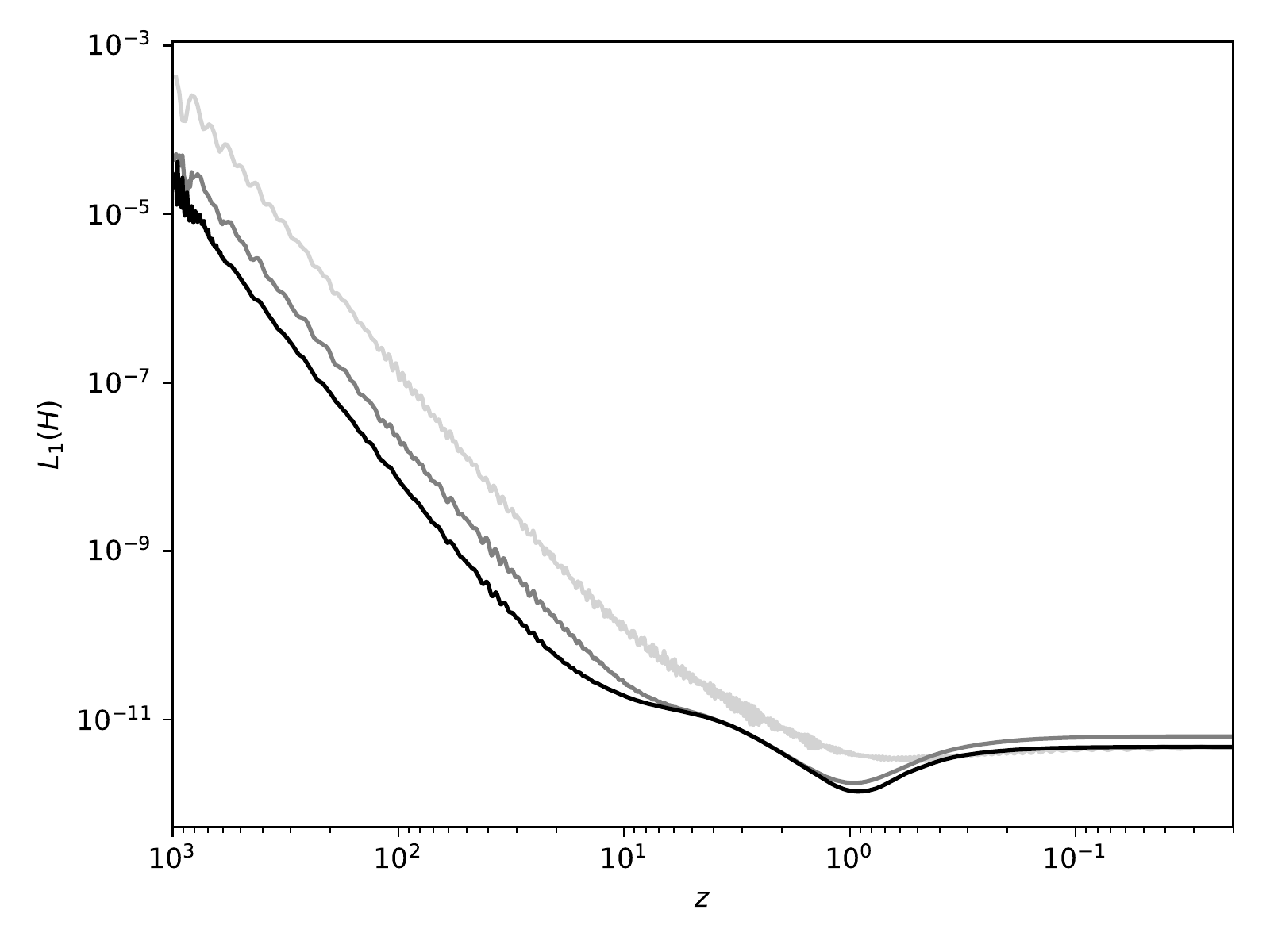} 
\includegraphics[width=0.495\columnwidth]{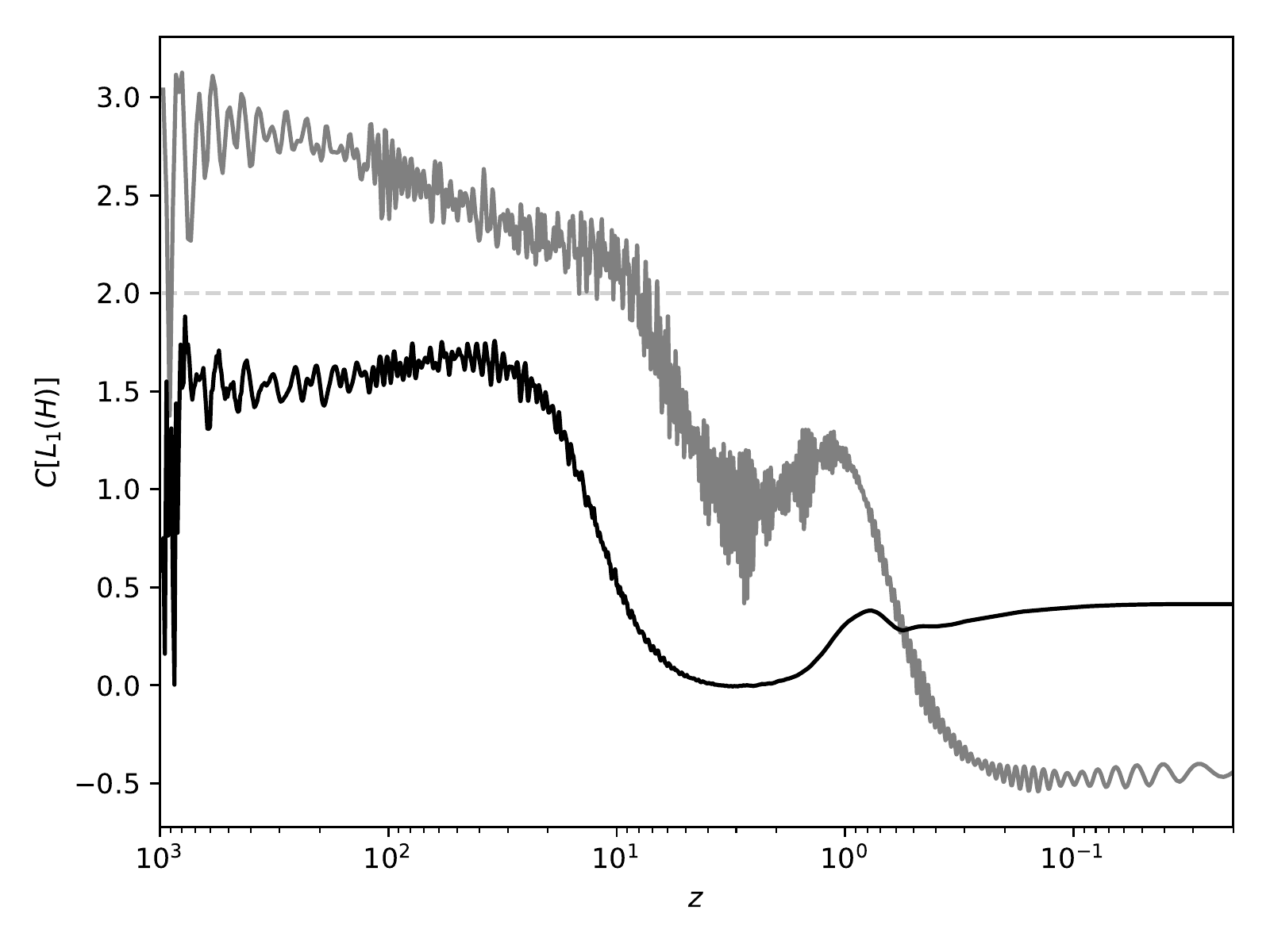} 
\caption{The $L_1$ measure of the absolute Hamiltonian constraint $H$ (left) and its converge ratio $C[L_1(H)]$ (right) as a function of redshift for the typical cosmology test with $L=256$. The left plot displays the three resolutions: $\De x = L/64$ (light gray), $\De x = L/128$ (gray) and $\De x = L/256$ (black), while the right plot displays the two resolution ratios $C_{L/128}[L_1(H)]$ (gray) and $C_{L/256}[L_1(H)]$ (black).}
\label{fig:cosmo_H}
\end{figure}

\begin{figure}[htbp]
\includegraphics[width=0.495\columnwidth]{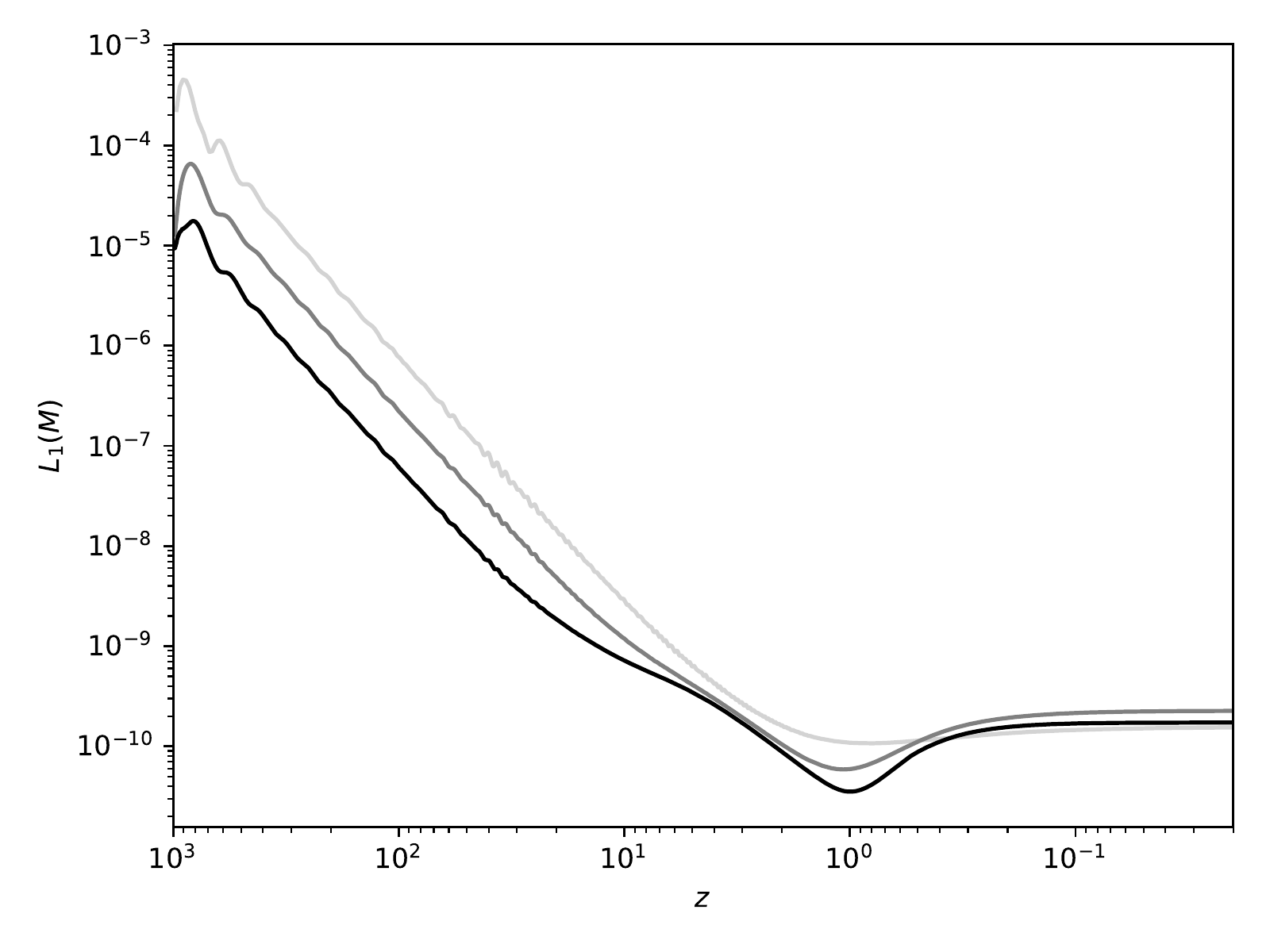} 
\includegraphics[width=0.495\columnwidth]{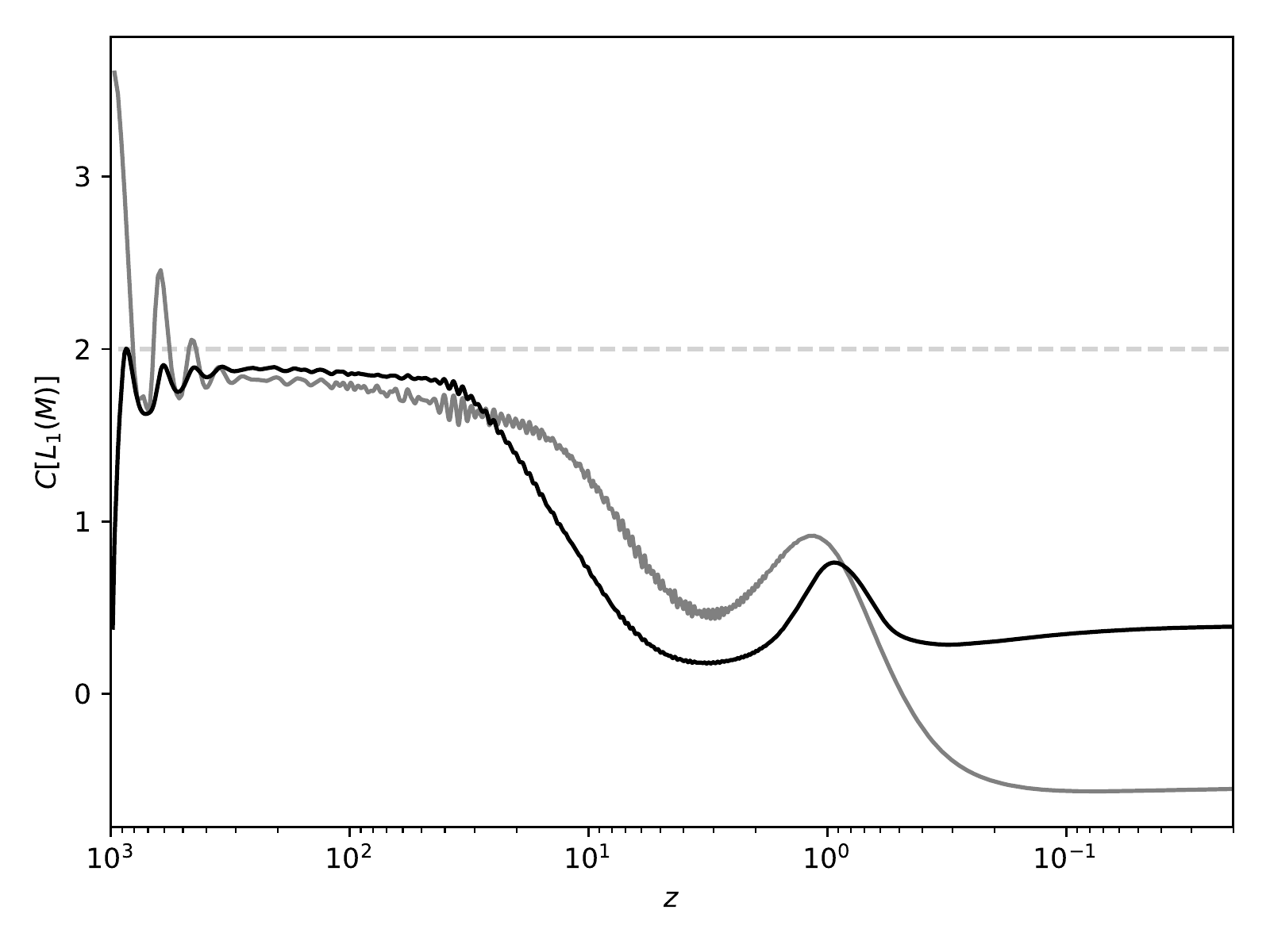} 
\caption{The $L_1$ measure of the absolute Hamiltonian constraint $M$ (left) and its converge ratio $C[L_1(M)]$ (right) as a function of redshift for the typical cosmology test with $L=256$. The left plot displays the three resolutions: $\De x = L/64$ (light gray), $\De x = L/128$ (gray) and $\De x = L/256$ (black), while the right plot displays the two resolution ratios $C_{L/128}[L_1(M)]$ (gray) and $C_{L/256}[L_1(M)]$ (black).}
\label{fig:cosmo_M}
\end{figure}

Let us now consider the simulation of smaller size $L = 256$. The analogues of figures \ref{fig:cosmo_mac_constr}, \ref{fig:cosmo_mac_H} and \ref{fig:cosmo_mac_M} are now figures \ref{fig:cosmo_constr}, \ref{fig:cosmo_H} and \ref{fig:cosmo_M}, respectively. At high redshift the behavior of the constraints is similar to the previous simulation, i.e. their amplitude and convergence is controlled. However, we observe two significant ``jumps", first at $z \sim 10$, after which convergence is no longer achieved at all, and then at $z \sim 1$. The latter is clearly due to the strong inhomogeneity that develops at these times and should be avoided by using adaptive mesh techniques. As for the first jump at $z \sim 10$, we observe that it coincides with the moment at which the particle number per cell develops spurious sharp variations in space. These are smoothed out when projected on the mesh to build the corresponding fields, but still strong enough to affect constraint violation. These features are  subsequently washed away by the formation of structure. We believe that this effect is due to our initial over-sampling of phase space, i.e. $N_{\rm ppc} = 27$. The problem is that we cannot lower this parameter, because then we under-sample the voids at late times and this leads to important constraint violation. It therefore seems that adaptive phase space resolution methods will cure this problem. A detailed investigation of this issue will be presented in another paper of this series. Note, also, that despite the aforementioned issue, we are able to control the relative constraints with an $L_1$ measure of at most $\sim 10^{-6}$ for the Hamiltonian and $\sim 10^{-2}$ for the momentum, all the way down to redshift zero, for all three resolutions and for both box sizes. Finally, note that the present test is a scalar linear multi-mode test at high enough redshits where evolution is linear. It is therefore interesting to compare with the results of the single linear mode test of the previous subsection. As already mentioned there, we see that here the relative constraints $H_r$ and $M_r$ converge with resolution (figures \ref{fig:cosmo_mac_constr} and \ref{fig:cosmo_constr}), contrary to the single mode case (figure \ref{fig:sm_de_Hr}). 

\begin{figure}[htbp]
\includegraphics[width=\columnwidth]{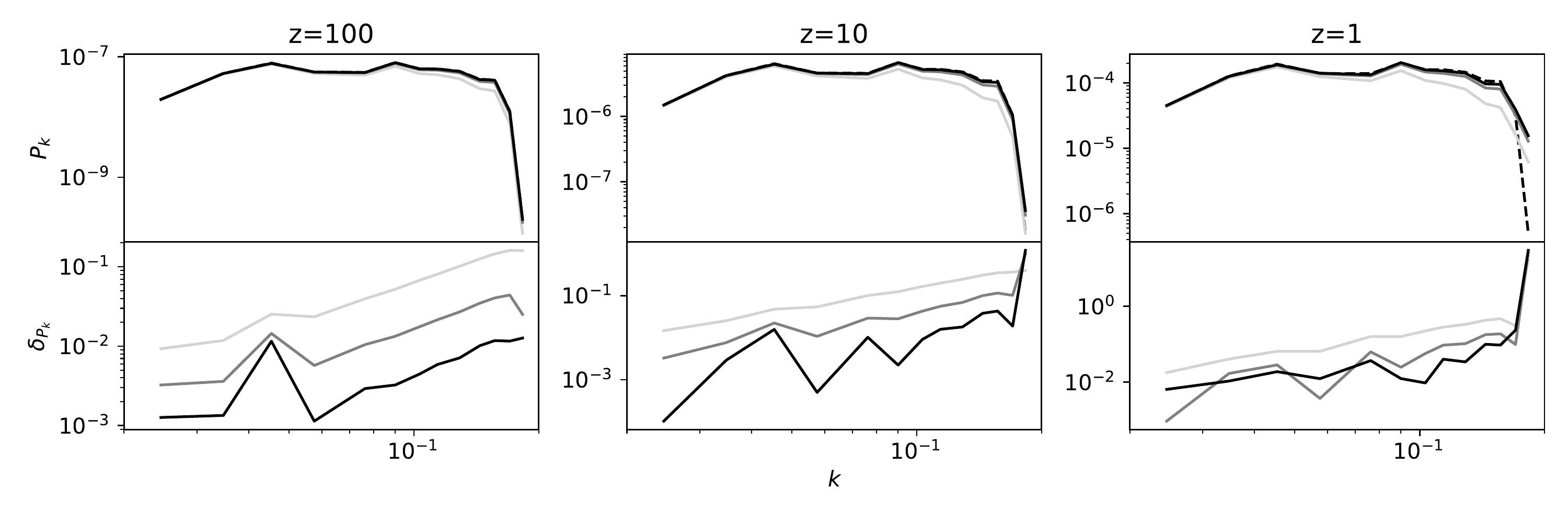} 
\caption{The power spectrum $P(t,k)$ (upper panels) and its relative deviation from the linear analytical solution $\de_{P_k}$ (lower panels) for the simulation with box size $L = 256$. The three panels from left to right display the redshifts $z \in \{ 100, 10, 1 \}$, respectively. Each panel contains the three resolutions $\De x = L/64$ (light gray), $\De x = L/128$ (gray) and $\De x = L/256$ (black), while the dashed line is the initial power spectrum that is propagated in time using the linear analytical growth function.}
\label{fig:cosmo_pks}
\end{figure}

Although controlling constraint violation is a necessary condition for an accurate resolution of the dynamics, it is not sufficient, and it can even be a misleading check when one uses a scheme that is precisely designed to dissipate that violation. We therefore now consider two more types of error for the $L = 256$ simulation. First, we verify that we are accurately resolving the matter power spectrum in the linear regime, for which we have the analytic solution \eqref{eq:Pkevol}. Numerically, it is computed by Fourier transforming the field $\de(t,\vec{x})$ and averaging its modulus squared over the angles
\beq
P_{\rm num.}(t,k) := \frac{1}{4\pi} \int \ed \Om_k\, \de(t,\vec{k})\, \de^*(t,\vec{k}) \, ,
\eeq 
that is approximated by averaging in $\De k = 1.28$ shells in practice. In figure \ref{fig:cosmo_pks} we plot this function of $k$ for the three redshift values $z = \{ 100, 10, 1 \}$ and also the relative difference with respect to the linear analytical solution
\beq
\de_{P_k}(t,k) := \frac{|P_{\rm num.}(t,k) - P_{\rm an.}(t,k)|}{P_{\rm an.}(t,k)} \, .
\eeq 
Here $P_{\rm an.}$ is constructed using \eqref{eq:Pkevol} where the initial $P_{\rm an.}(0,k)$ is the one of the initial state of the simulation. Each plot contains all three resolutions and also the analytical one (dashed) and we observe a deviation of the order of $\sim 10^{-2}$ at worst for the best resolution. The growth of error with increasing $k$ is to be expected, since the linear approximation becomes less and less valid.

\begin{figure}[htbp]
\includegraphics[width=0.495\columnwidth]{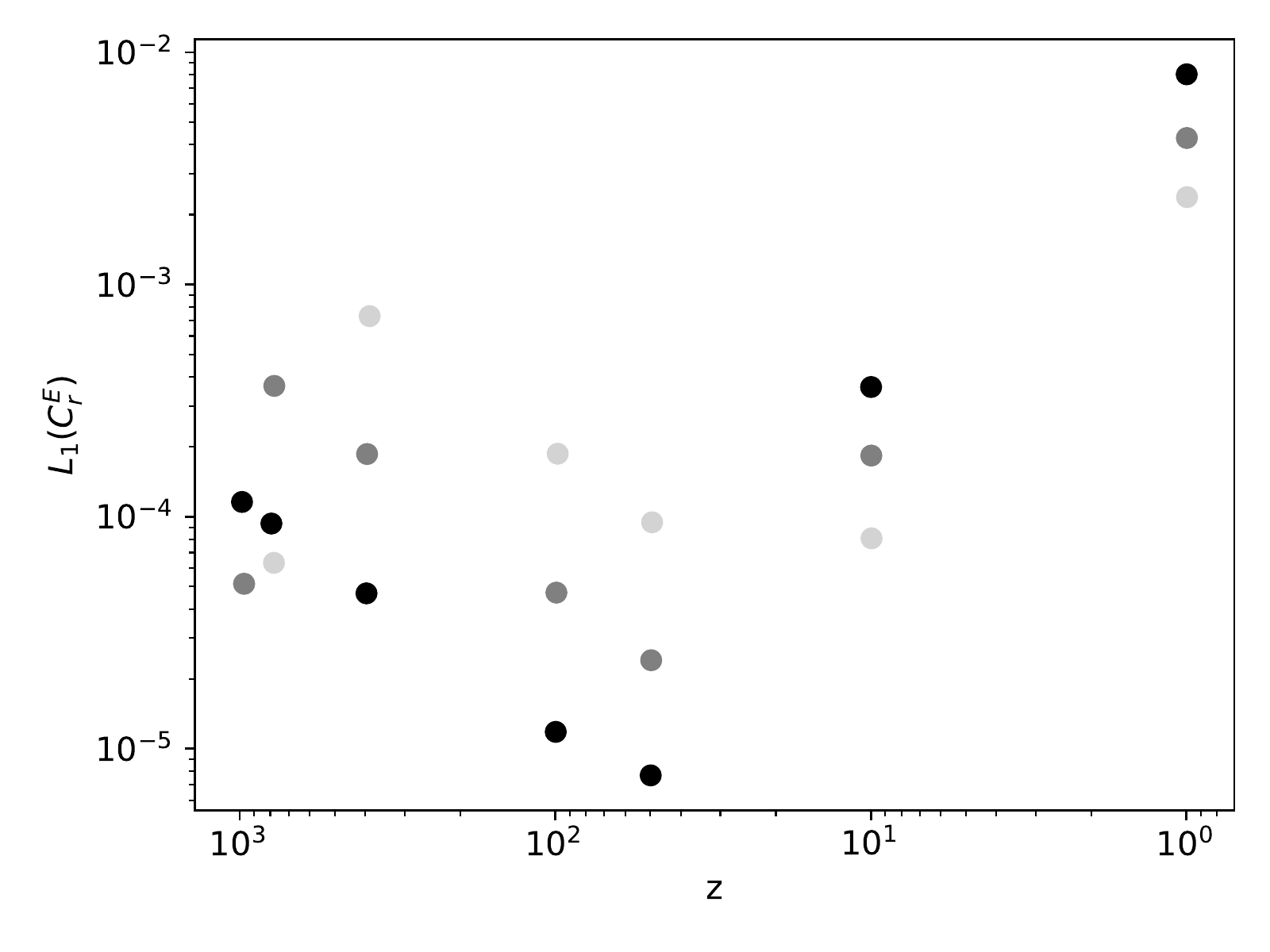} 
\includegraphics[width=0.495\columnwidth]{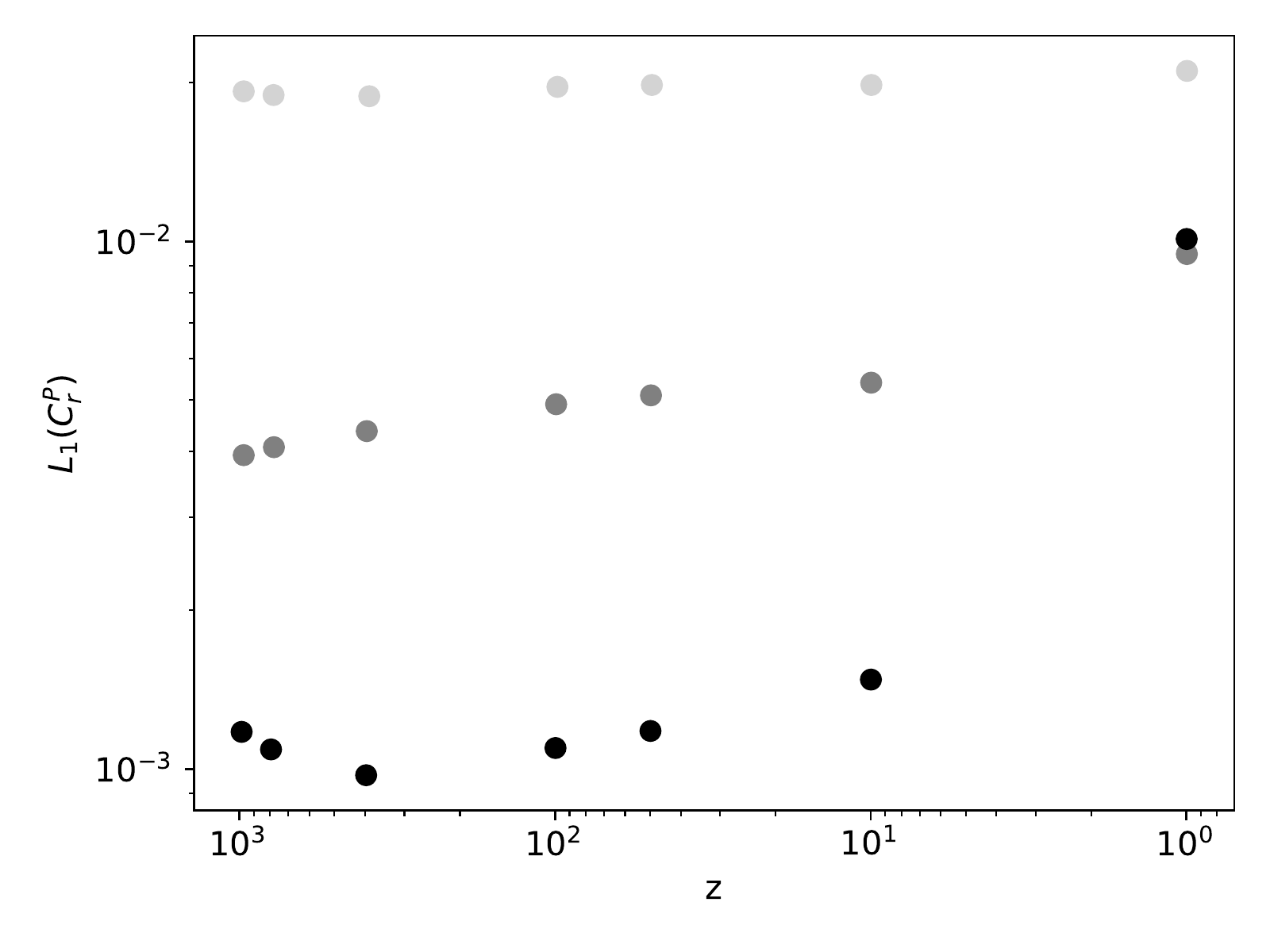} 
\caption{The $L_1$ measure of the relative energy $C_r^E$ (left) and momentum $C_r^P$ (right) conservation violation for the redshift values $z = \{ 1000, 800, 400, 100, 50, 10, 1 \}$ for the typical cosmology test with $L=256$ and for the three resolutions: $\De x = L/64$ (light gray), $\De x = L/128$ (gray) and $\De x = L/256$ (black).}
\label{fig:ec_cosmo}
\end{figure}

Second, we also consider whether the violation of energy-momentum conservation is under control. Contrary to the previous check, here we can measure an error at the fully non-linear level, so we do not need to restrict our attention to the large scale Fourier modes only. In terms of the present variables, the conservation equations $\na_{\mu} T^{\mu\nu} = 0$ read
 \bea
C^E & := & \dot{E} - \al K E - \al \ch \ti{\Ga}^i P_i + \ti{\ga}^{ij} \[ \al \ch \pa_i P_j + 2 \ch P_i \pa_j \al - \frac{1}{2}\, \al P_i \pa_j \ch - \frac{1}{3}\, \al \ch K S_{ij} \] - \al \ch \ti{\ga}^{ik} \ti{\ga}^{jl} \ti{A}_{ij} S_{kl} \nn \\
 & = & 0 \, , \label{eq:CE}  \\
C^P_i & := & \dot{P}_i - \al K P_i + E \pa_i \al - \al \ch \ti{\Ga}^j S_{ij}  \label{eq:CP} \\
 & & +\, \ti{\ga}^{jk} \[ \al \ch \( \pa_j S_{ik} - S_{lk} \ti{\Ga}^l_{ij} \) + \ch S_{ij} \pa_k \al + \frac{1}{2}\, \al \( S_{jk} \pa_i \ch - S_{ij} \pa_k \ch \) \] = 0 \, . \nn  
\eea
and the time-derivative is computed using a three-point stencil, involving three successive loop time-steps, i.e. separated by $N_s \De t$ (see subsection \ref{sec:timeint}). As with the constraint equations, here too the relevant quantities are the relative ones with respect to the typical magnitude of the involved terms
\beq \label{eq:CErCPr}
C^E_r := \frac{|C^E|}{\sqrt{\sum_n T^E_n T^E_n}} \, , \hspace{1cm}  C^P_r := \frac{\sqrt{C^P_i C^P_i}}{\sqrt{\sum_n \ga^{ij} T^P_{n,i} T^P_{n,j}}} \, ,
\eeq
where $T^E_n$ and $T^P_{n,i}$ denote the $n$-th term appearing on the right-hand side of \eqref{eq:CE} and \eqref{eq:CP}, respectively. In figure \ref{fig:ec_cosmo} we show a rough plot of the time evolution of the $L_1$ norm of these two quantities, by displaying in particular the value for the redshifts $z = \{ 1000, 800, 400, 100, 50, 10, 1 \}$. We observe an average relative error of $\sim 10^{-2}$ for both $C^E$ and $C_i^P$ for all considered resolutions and redshifts. Finally, as in the case of the constraints, convergence is generically lost at small redshifts. 

\section{Conclusion}  \label{sec:conclusion}

In this paper we have proposed a numerical prescription for the time-evolution of $N$-body NR simulations in cosmology. We have shown that it passes the robustness and scalar linear mode tests around the FLRW solution. We then considered the evolution of typical cosmological initial conditions and showed that our code follows the linear part of the power spectrum accurately and controls well the violation of the constraints and energy-momentum conservation. However, convergence is not achieved for low redshift and small scale simulations. Nevertheless, we have argued that this problem is related to the fact that we work with a Cartesian mesh and with fixed number of particles. It could therefore probably be resolved by considering adaptive resolution methods both for the mesh and the phase space samplers.

\appendix

\section{Derivation of the particle equations} \label{app:partder}

The action for a set of $N$ minimally coupled particles of mass $m$ is given by
\beq
S_N = - m \sum_{a=1}^N \int \ed \la\, \sqrt{-g_{\mu\nu}(x_a)\, \dot{x}_a^{\mu} \dot{x}_a^{\nu}} \, , 
\eeq
where $\la$ is an arbitrary parameter and here the dot denotes the derivative with respect to it. We fix the $\la$-reparametrization gauge by requiring that $\la$ coincides with the space-time coordinate $x^0 \equiv t$
\beq
x^0(\la) = \la \, , \hspace{1cm} \Rightarrow \hspace{1cm} x^i(\la) \to x^i(t) \, ,
\eeq
so now the dot coincides with $\pa_t$. Note that this is not the proper time of the particle, unless we also choose to fix $\al = 1$, which is not the case here. We next express the action in terms of the ADM variables
\beq
S_N = - \sum_{a=1}^N m_a \int \ed t \, \sqrt{\al^2(\vec{x}_a) - \ga_{ij}(\vec{x}_a) \[ \dot{x}_a^i + \be^i(\vec{x}_a) \] \[ \dot{x}_a^j + \be^j(\vec{x}_a) \]} \, , 
\eeq
compute the conjugate momenta of $x^i_a$
\beq
p^a_i := \frac{\pa L}{\pa \dot{x}^i_a} = \frac{m_a \ga_{ij}(\vec{x}_a) \[ \dot{x}_a^j + \be^j(\vec{x}_a) \]}{\sqrt{\al^2(\vec{x}_a) - \ga_{ij}(\vec{x}_a) \[ \dot{x}_a^i + \be^i(\vec{x}_a) \] \[ \dot{x}_a^j + \be^j(\vec{x}_a) \]}} \, ,
\eeq
whose inverse relation reads
\beq
\dot{x}_a^i + \be^i(\vec{x}_a) = (\al \ga^{ij})(\vec{x}_a)\, \frac{p_j^a}{E_a} \, , \hspace{1cm} E_a := \sqrt{m_a^2 + \ga^{ij}(\vec{x}_a)\, p_i^a p_j^a} \, ,
\eeq
and Legendre transform with respect to $\dot{x}_a^i$ to get the canonical action
\beq
S_N = \int \ed t \[ p_i^a \dot{x}^i_a - \al(\vec{x}_a)\, E_a + \be^i(\vec{x}_a) \, p_i^a \]  \, .
\eeq
The canonical energy-momentum components are therefore
\bea
E & = & \sum_{a=1}^N E_a \, \frac{\de^{(3)} (\vec{x} - \vec{x}_a)}{\sqrt{\ga}} \, , \nn \\
P_i & = & \sum_{a=1}^N p_i^a \, \frac{\de^{(3)} (\vec{x} - \vec{x}_a)}{\sqrt{\ga}} \, , \label{eq:EPSder} \\
S_{ij} & = & \sum_{a=1}^N \frac{p_i^a p_j^a}{E_a} \, \frac{\de^{(3)} (\vec{x} - \vec{x}_a)}{\sqrt{\ga}} \, . \nn
\eea
and the equations of motion are
\bea
\dot{x}_a^i = - \be^i + \al \,  \frac{\ga^{ij} p_j^a}{E_a} \, , \hspace{1cm} \dot{p}^a_i = - E_a \pa_i \al + p_j^a \pa_i \be^j + \al \, \frac{\ga^{jl} \Ga^k_{il} p_j^a p^a_k}{E_a} \, , 
\eea
where it is understood that the above fields are evaluated at $\vec{x}_a$ and $\Ga^k_{ij}$ are the Christoffel symbols of $\ga_{ij}$. Expressing these equations in terms of the conformally decomposed variables one obtains \eqref{eq:partevol}, while the ``discretization" of the Dirac delta in \eqref{eq:EPSder} yields \eqref{eq:EPS}.

\section{Linear perturbation theory equations} \label{app:perts}

Here we work with the analytical solution \eqref{eq:alsol} and model matter as a pressureless perfect fluid, which is a valid description in the regime where linear perturbation theory applies since the velocity field is smooth. We introduce perturbations around the FLRW solution given in \eqref{eq:FLRW} 
\bea
Q & = & 1 + \psi \, , \label{eq:Q} \\
\chi & = & a^{-2} \( 1 + 2 \vph \) \, , \\
\al & = & a \( 1 + \psi - \vph \) \, , \\
K & = & -3 a^{-1} \cH \( 1 + h \) \, , \\
\ti{\ga}_{ij} & = & \de_{ij} + 2 \( \pa_i \pa_j - \frac{1}{3}\, \de_{ij} \pa^2 \) \ti{\vph}  \, , \\
\ti{A}_{ij} & = & - a^{-1} \cH \( \pa_i \pa_j - \frac{1}{3}\, \de_{ij} \pa^2 \) \ti{h}  \, , \\
E & = & 3 a^{-2} \cH^2 \( 1 + \de \) \, , \label{eq:deE} \\
P_i & = & 3 a^{-1} \cH^2 \pa_i v \, .   \label{eq:dePi} 
\eea
where $\psi$ is time-independent and its profile amounts to the residual gauge choice of initial conditions for $\al$. Note also that $v$ is a comoving velocity. Going to Fourier space, the linearized constraint equations read
\beq \label{eq:Hlin} 
\de = 2 h - \frac{2}{3} \frac{k^2}{\cH^2}\, \vph + \frac{2}{9} \frac{k^4}{\cH^2} \, \ti{\vph} \, ,  \hspace{1cm} v = \frac{2}{3\cH} \( h + \frac{1}{3}\, k^2 \ti{h} \) \, , 
\eeq
while the evolution equations yield
\bea
\dot{\vph} & = & \cH \[ \vph - \psi - h \] \, , \\
\dot{\ti{\vph}} & = & \cH \ti{h} \, ,  \label{eq:vphdot} \\
\dot{h} & = & \cH \[ \frac{3}{2} \( \vph - \psi \) - \frac{9}{2}\, h + \frac{k^2}{\cH^2} \( \frac{5}{3}\, \vph - \frac{1}{3}\, \psi - \frac{4}{9} \, k^2 \ti{\vph} \) + \frac{3}{2}\, \de \] \, , \\
\dot{\ti{h}} & = & \cH \[ - \frac{3}{2}\, \ti{h} + \frac{1}{3}\, \frac{k^2}{\cH^2}\, \ti{\vph} + \frac{1}{\cH^2} \( \psi - 2 \vph \)  \] \, .
\eea
Using \eqref{eq:Hlin} to eliminate $\de$, the scalar evolution equations can be written in closed second-order form
\bea
\ddot{\vph} + \cH \dot{\vph} + \frac{2}{3} \, k^2 \vph & = & \frac{1}{3}\, k^2 \( \psi + \frac{1}{3}\, k^2 \ti{\vph} \) \, , \label{eq:vphevol} \\
\ddot{\ti{\vph}} + 2 \cH \dot{\ti{\vph}} - \frac{1}{3} \, k^2\, \ti{\vph} & = & \psi - 2 \vph \, . \label{eq:noanstress}
\eea
In terms of these variables the scalar Bardeen potentials are
\beq \label{eq:Bardeen}
\Psi := \psi - \vph - \ddot{\ti{\vph}} - \cH \dot{\ti{\vph}} \, , \hspace{1cm} \Phi := \vph - \frac{1}{3}\, k^2 \ti{\vph} + \cH \dot{\ti{\vph}} \, , 
\eeq
while the gauge-invariant matter quantities are
\beq \label{eq:GIconstr}
\de_{\star} := \de - 3 \cH v \, , \hspace{1cm} v_{\star} := v + \dot{\ti{\vph}} \, .
\eeq
The scalar constraint equations can now be expressed as
\beq \label{eq:linconstr}
k^2 \Phi = -\frac{3}{2}\, \cH^2 \de_{\star} \, , \hspace{1cm} \dot{\Phi} + \cH \Psi = - \frac{3}{2}\, \cH^2 v_{\star} \, ,
\eeq
while \eqref{eq:noanstress} is nothing but the absence of anisotropic stress $\Psi = \Phi$. Let us now fix the residual gauge freedom, starting with
\beq \label{eq:psi}
\psi(\vec{x}) = 2 \vph(0, \vec{x}) \, ,
\eeq
so that the linearized gauge conditions read
\beq
\frac{\de \al}{\bar{\al}} + \vph = 2 \vph(0) \, , \hspace{1cm} \be^i = 0 \, .
\eeq
These are preserved under a gauge transformation with generating vector $\xi^{\mu} = \( \xi^t, \de^{ij} \pa_j \xi \)$ obeying
\beq \label{eq:xi}
\dot{\xi}^t = -\frac{1}{3}\, k^2 \xi - 2\[ \cH \xi^t - \frac{1}{3}\, k^2 \xi \](0) \, , \hspace{1cm} \dot{\xi} = \xi^t \, .
\eeq
This is a first-order system for $\xi^t$ and $\xi$, so their initial conditions are free to choose. To use this freedom we note that, under a general gauge transformation
\bea
\de_{\xi} \ti{\vph} & = & - \xi \, , \hspace{1cm} \de_{\xi} \ti{h} = - \cH^{-1} \dot{\xi} \, ,
\eea
where we used \eqref{eq:vphdot} for the latter. Thus, under a residual gauge transformation at initial time $t = 0$, using \eqref{eq:xi}, we find
\bea
\de_{\xi} \ti{\vph}(0) = - \xi(0) \, , \hspace{1cm} \de_{\xi} \ti{h}(0) = - \[ \cH^{-1} \xi^t \] (0) \, ,
\eea 
which allows us to set 
\beq
\ti{\vph}(0,\vec{x}) = \ti{h}(0,\vec{x}) = 0 \, .
\eeq
The advantage of this gauge is that, under the assumption of the Zel'dovich condition at $t=0$
\beq \label{eq:Zeldovich}
\dot{\Phi}(0) = 0 \hspace{1cm} \Rightarrow \hspace{1cm} \vph(0) + h(0) = 0 \, ,
\eeq
the gravitational fields remain constant in time 
\beq
\vph(t,\vec{x}) = \vph(0,\vec{x}) \, , \hspace{1cm} \ti{\vph}(t,\vec{x}) = 0 \, , \hspace{1cm} h(t,\vec{x}) = - \vph(0,\vec{x}) \, , \hspace{1cm} \ti{h}(t,\vec{x}) = 0 \, , 
\eeq
and the corresponding line-element takes the conformal Newtonian form
\beq 
\ed s^2 = a^2 \[ - \( 1 + 2 \vph \) \ed t^2 + \( 1 - 2 \vph \) \ed \vec{x}^2 \] \, .
\eeq
We are therefore in the conformal Newtonian gauge in the scalar sector, but only with our choice of initial conditions \eqref{eq:Zeldovich} and evolution equations, i.e. the fact that the considered theory is GR. The only evolving quantities are the matter density and velocity, because of the $\cH$ factors in \eqref{eq:Hlin}, and now read
\beq \label{eq:devevol}
\de(t,\vec{x}) = - 2 \[ 1 - \frac{1}{3} \frac{\pa^2}{\cH^2(t)} \] \vph(0,\vec{x})  \, , \hspace{1cm} v(t,\vec{x}) = - \frac{2}{3\cH(t)} \,\vph(0,\vec{x}) \, .
\eeq
With the first of these equations we can then infer the evolution of the matter power spectrum
\beq \label{eq:Pkevol}
P(t,k) = \frac{3 + k^2 / \cH^2(t)}{3 + k^2/\cH^2(0)}\, P(0,k) \, .
\eeq

\acknowledgments

We are grateful to Hayley Macpherson, Joachim Stadel and Romain Teyssier for useful discussions and especially to Martin Kunz for helpful support. This work has been supported by an Advanced Postdoc.Mobility grant of the Swiss National Science Foundation for DD, by a Consolidator Grant of the European Research Council (ERC-2015-CoG grant 680886) for YD and EM and by a grant from the Swiss National Supercomputing Centre (CSCS) under project ID s751. The simulations have been performed on the Baobab cluster of the University of Geneva and on Piz Daint of the CSCS.

\bibliographystyle{JHEP}
\bibliography{cosmo_NR}

\end{document}